\documentclass[12pt]{article}
\pdfoutput=1

\usepackage{amsmath,amssymb,amsfonts,graphicx,amsfonts}
\usepackage{textcomp,mathcomp}
\usepackage{color,xcolor}
\usepackage[hidelinks]{hyperref}
\usepackage{ascmac}
\usepackage{physics}

\allowdisplaybreaks[4]

\date{}
\begin{document}
\begin{flushright}
\today\\
\end{flushright}

\vspace{0.1cm}

\begin{center}

  {\Large Wilson Loops and Random Matrices}\\

\end{center}
\vspace{0.1cm}
\vspace{0.1cm}
\begin{center}

Georg Bergner$^{a}$, Vaibhav Gautam$^{b,c}$, Masanori Hanada$^{b}$, Jack Holden$^{d,b}$

\end{center}
\vspace{0.3cm}

\begin{center}

{\small

$^a$ University of Jena, Institute for Theoretical Physics\\
Max-Wien-Platz 1, D-07743 Jena, Germany\\
\vspace{3mm}
$^b$ School of Mathematical Sciences, Queen Mary University of London\\
Mile End Road, London, E1 4NS, United Kingdom\\
\vspace{3mm}
$^c$ 
Department of Mathematics, University of Surrey\\
Guildford, Surrey, GU2 7XH, United Kingdom\\
\vspace{3mm}
$^d$ Yau Mathematical Sciences Center, Tsinghua University, Beijing, 100084, China\\
\vspace{3mm}
}
\end{center}

\vspace{1.5cm}

\begin{center}
  {\bf Abstract}
\end{center}

Linear confinement with Casimir scaling of the string tension in confining gauge theories is a consequence of a certain property of the Polyakov loop related to random matrices. This mechanism does not depend on the details of the theories (neither the gauge group nor dimensions) and explains approximate Casimir scaling below string-breaking length. In this paper, we study 3d SU(2) pure Yang-Mills theory numerically and find the same random-matrix behavior for rectangular Wilson loops. We conjecture that this is a universal feature of strongly coupled confining gauge theories.

\newpage
\tableofcontents

%%%%%%%%%%%%
%%%%%%%%%%%%
\section{Introduction}\label{sec: Intro}
\hspace{0.51cm}
%%%%%%%%%%%%
%%%%%%%%%%%%
In this paper, we demonstrate an intriguing property of the rectangular Wilson loop described by random matrices for three-dimensional SU(2) pure Yang-Mills theory. We conjecture that this is a universal feature of strongly coupled confining gauge theories. 

The Wilson loop has been studied for decades as a fundamental observable in gauge theory. When confinement takes place, a large Wilson loop exhibits an area law. Specifically, in the Euclidean path-integral formalism, a rectangular loop of size $L_t$ along the imaginary time direction and $L_s$ along the spatial direction, which we denote by $W(L_t\times L_s)$, behaves as 
\begin{align}
\langle W(L_t\times L_s)\rangle
\sim
e^{-\sigma^2L_tL_s}\, , 
\end{align}
where $\sigma^2$ is the string tension. (Strictly speaking, this is the case below the string-breaking length scale and above a certain short-distance scale.) 

Usually, Wilson loops in the fundamental representation of the gauge group are studied. In general, however, Wilson loops in any representations are well-defined and provide us with detailed information about the gauge theory. An intriguing pattern observed in various gauge theories is the Casimir scaling of string tension~\cite{Ambjorn:1984mb,Ambjorn:1984dp,DelDebbio:1995gc,Bali:2000un,Deldar:1999vi}, i.e., for each irreducible representation r, the string tension is proportional to its quadratic Casimir $C_{\rm r}$: 
\begin{align}
\langle W^{\rm (r)}(L_t\times L_s)\rangle
\sim
e^{-\sigma_0^2C_{\rm r}L_tL_s}\, . 
\end{align}
Here, $\sigma_0^2$ is a number that is independent of representation and the quadratic Casimir $C_{\rm r}$ is defined using generators in representation r, denoted by $T^{\rm (r)}$, as 
\begin{align}
\sum_\alpha \left(T^{\rm (r)}_\alpha\right)^2 = 4C_{\rm r}\cdot\textbf{1}\, ,
\end{align}
where the generators of the Lie algebra $T_{\alpha}$ ($\alpha=1,2,\cdots,N^2-1$) are normalized for the fundamental representation as 
\begin{align}
\mathrm{Tr}(T_\alpha T_\beta)=2\delta_{\alpha\beta}\, .
\end{align}
 As an example, $C_{{\rm spin}\mathchar`-j}=j(j+1)$ for spin-$j$ representation of SU(2), while for fundamental, adjoint, rank-2 symmetric, and rank-3 symmetric representations of SU(3), $C_{\rm fund}=\frac{4}{3}$, $C_{\rm adj}=3$, $C_{\rm 2\mathchar`-sym}=\frac{10}{3}$, and $C_{\rm 3\mathchar`-sym}=6$, respectively. 
The Casimir scaling holds approximately and at intermediate length scale, i.e., at sufficiently long distance so that the area law can hold, and at the same time, at sufficiently short distances so that string breaking does not take place.

Recently, it was pointed out that Polyakov loops have a certain kind of randomness~\cite{Hanada:2023krw,Hanada:2023rlk} and Casimir scaling is a consequence of such randomness~\cite{Bergner:2023rpw}.\footnote{Earlier references that considered a random walk on the group manifold and its potential connection to confinement include Refs.~\cite{Brzoska:2004pi,Arcioni:2005iq,Buividovich:2006yj,Buividovich:2007xh}.
While these references introduced a random walk based on phenomenological observation from lattice simulations, Refs.~\cite{Hanada:2023krw,Hanada:2023rlk,Bergner:2023rpw} identified the microscopic mechanism behind the random walk, and furthermore, differences from the vanilla random walk (see Sec.~\ref{sec:review}) were taken into account. 
}  
Furthermore, it was conjectured that Wilson loops have the same kind of randomness. In this paper, we give numerical evidence for this conjecture by taking three-dimensional SU(2) pure Yang-Mills theory as a concrete example, using Wilson's plaquette action as a regularization.

Let us consider a rectangular Wilson loop of size $L_t\times L_s$ in the confined phase. To have well-defined behavior in the continuum limit, including at short distances, we use a loop made of renormalized gauge fields. Specifically, we use the gradient flow~\cite{Luscher:2010iy,Datta:2015bzm,Petreczky:2015yta} for renormalization. It leads to a smearing of gauge fields that does not affect long-distance behavior, which we are ultimately interested in. 

\begin{figure}[htbp]
\begin{center}
\scalebox{0.25}{
\includegraphics{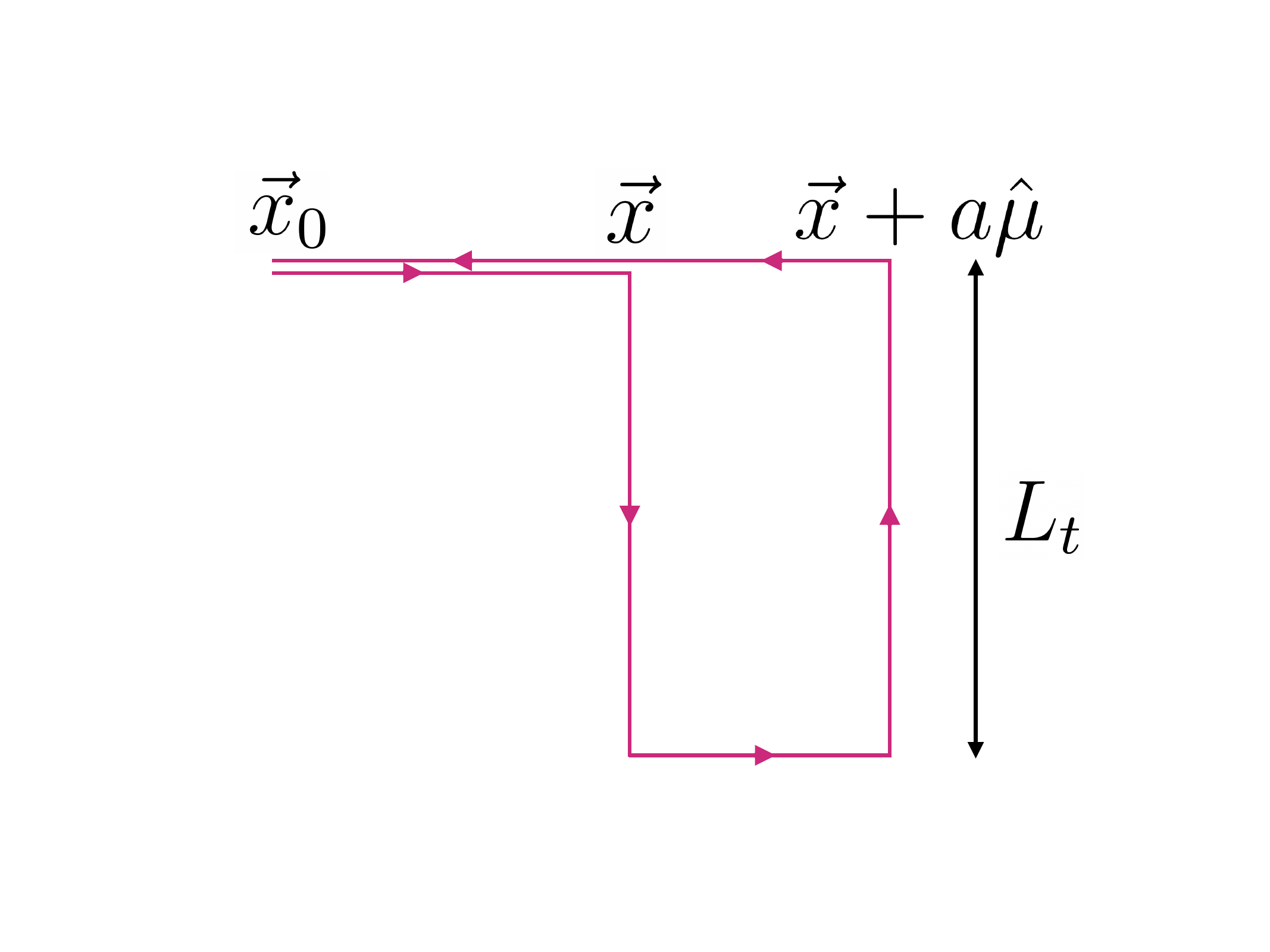}}
\scalebox{0.4}{
\includegraphics{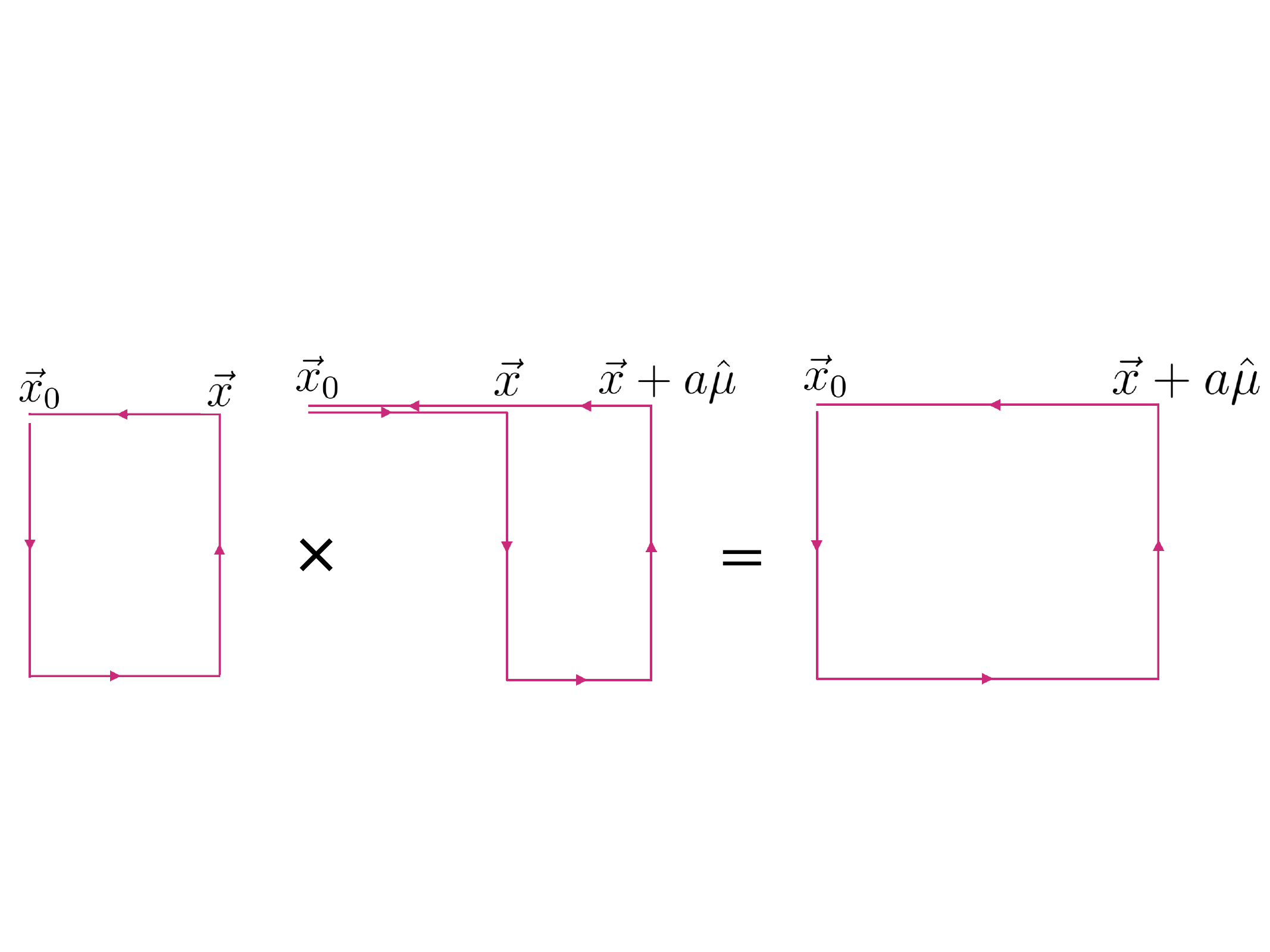}}
\end{center}
\caption{
[Top]
A Wilson line that can be used to build a rectangular Wilson loop of the size $L_t$ for the time direction and $L$ for the spatial direction. 
We call it $W_{L_t}(\vec{x},\vec{x}+a\hat{\mu})$. 
[Bottom] How $W_{L_t}(\vec{x},\vec{x}+a\hat{\mu})$ can be used to build a rectangular Wilson loop. 
}\label{fig:basic-idea}
\end{figure}

We consider Wilson lines\footnote{In our notation, we denote Wilson loops as the trace of closed Wilson lines.} $W_{L_t}(\vec{x},\vec{x}+a\hat{\mu})$ along special thin closed contours with both ends connected to a given point $\vec{x}_0$ on the lattice as defined in Fig.~\ref{fig:basic-idea}(top). These lines are elementary building blocks and larger contours with extend $L_t$ in time direction can be generated from products of them. For example, closed lines of size $L_t\times L_s$ are obtained by multiplying $L_s/a$ of these lines, see Fig.~\ref{fig:basic-idea}(bottom),
\begin{align}
W_{L_t}(\vec{x}_0,\vec{x}_0+a\hat{\mu})
\times
W_{L_t}(\vec{x}_0+a\hat{\mu},\vec{x}_0+2a\hat{\mu})
\times
\cdots
\times
W_{L_t}(\vec{x}_0+(L_s-a)\hat{\mu},\vec{x}_0+L_s\hat{\mu})\, . 
\end{align} 
The reason for introducing these lines, which we call thin lines in the following, is to formulate the theory in terms of a map from the lattice to the group manifold at a single reference lattice point $\vec{x}_0$.

Our basic conjecture is that such thin loops have a certain kind of randomness up to small corrections. Specifically,
\begin{align}
W_{L_t}(\vec{x},\vec{x}+a\hat{\mu})=e^{i\sqrt{L_t}\epsilon X_{\mu,\vec{x}}}\, ,
\label{eq:WX}
\end{align}
where $X_{\mu,\vec{x}}$ is a random matrix with variance of order one, and $\epsilon$ is a small number that decreases with lattice spacing $a$. $X_{\mu,\vec{x}}$ changes gradually with $\vec{x}$ so that $X_{\mu,\vec{x}}$ becomes continuous in the continuum limit. Numerically, we find in this work that $X_{\mu,\vec{x}}$ is \textit{Gaussian} random for three-dimensional SU(2) pure Yang-Mills. 
%For sufficiently small lattice spacing, $\epsilon$ is proportional to $a$. 
In the continuum limit, 
\begin{align}
W_{L_t}(\vec{x},\vec{x}+L\hat{\mu})=e^{i\sqrt{L_t}\int_0^L dL'v(L')}\, , 
\label{eq:def_ov_v}
\end{align}
where $v(L')$ is a slowly varying velocity that describes a random walk on the group manifold starting at the identity. 

In two-dimensional pure Yang-Mills theory, we can show for bare Wilson loops that $\epsilon\sim\sqrt{a}$, and $X_{\mu,\vec{x}}$ becomes uncorrelated at a distance of one lattice unit. In this case, we can prove that $X_{\mu,\vec{x}}$ is Gaussian random. In higher dimensions, renormalization is needed to have a well-defined continuum limit of $X_{\mu,\vec{x}}$. For 3d and 4d, the natural scaling is $\epsilon\sim a$. 

\begin{figure}[htbp]
\begin{center}
\scalebox{0.2}{
\includegraphics{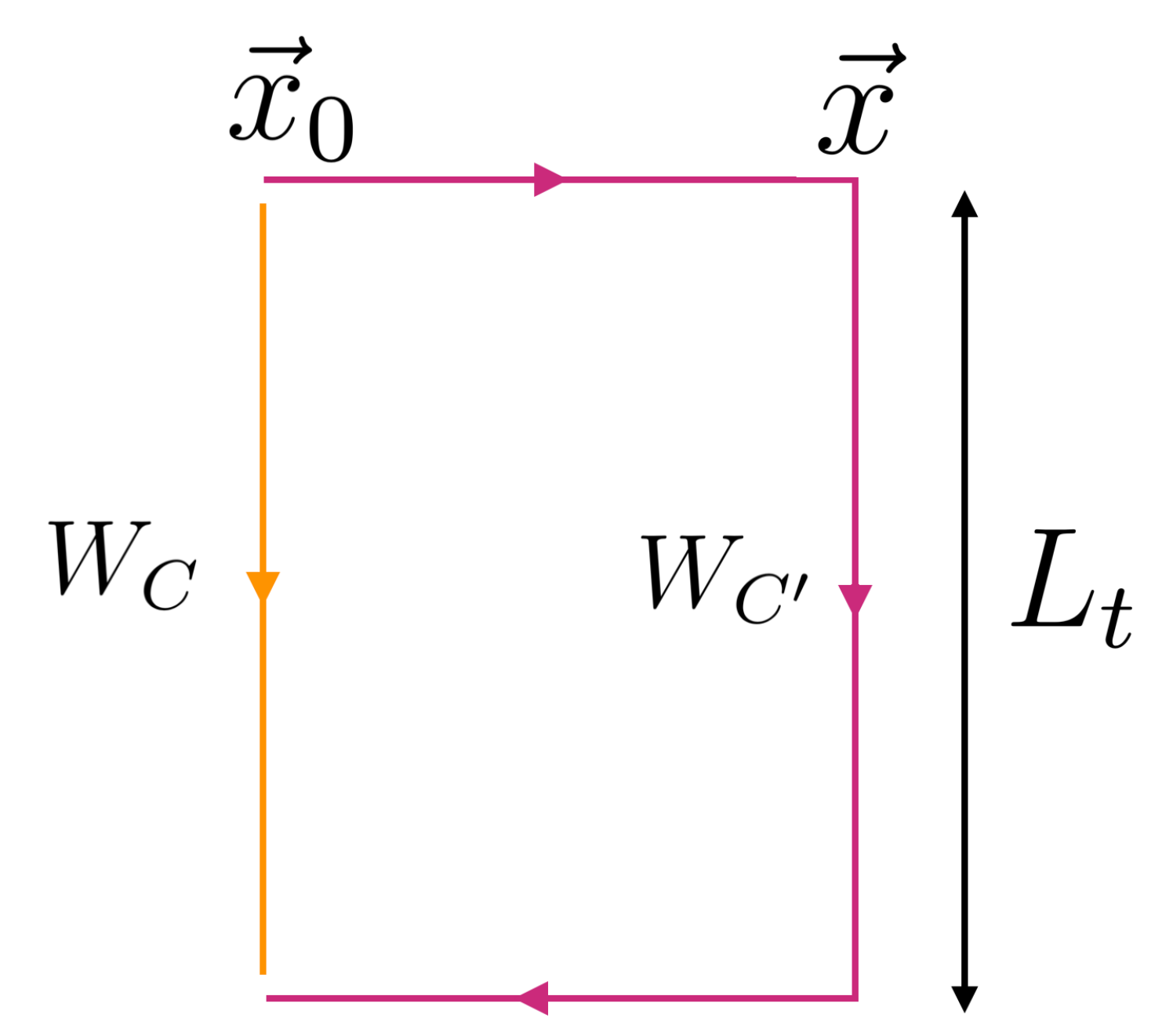}}
\end{center}
\caption{
Two open Wilson lines that can be used to form a rectangular Wilson loop.
$W_{C'}$ random walks on the SU($N$) group manifold as a function of $\vec{x}$, up to exponentially small corrections with respect to $L_t$. 
}\label{fig:Wilson-line-random-walk}
\end{figure}

We can rephrase this random-walk nature in terms of the open Wilson line. We can multiplyan open Wilson line $W_C$, which has a contour as depicted in Fig.~\ref{fig:Wilson-line-random-walk}, by thin Wilson lines to and get another open Wilson line $W_{C'}$. Then, as $\vec{x}$ moves further away from $\vec{x}_0$, $W_{C'}$ exhibits a random walk on the group manifold. 

When $L_t=\beta=T^{-1}$, where $T$ is temperature and $\beta$ is the circumference of the imaginary time direction, we can use $W_{L_t=\beta}$ to compute the two-point function of Polyakov loops~\cite{Bergner:2023rpw}. The Polyakov loops capture the symmetry of quantum states in the Hamiltonian formulation~\cite{Hanada:2020uvt}, and in the confined phase, they have a slowly varying Haar random property~\cite{Hanada:2023krw,Hanada:2023rlk} that leads to the randomness of $W_{L_t=\beta}$ mentioned above. We will provide a short review of the origin of the slowly varying Haar random property in Appendix~\ref{sec:Hanada-Watanabe-review}. 

For $L_t<\beta$, it is natural to expect a similar property, since although we lack a derivation based on the Hamiltonian formulation, there are obvious similarities to the path integral formulation. The present paper aims to confirm this property numerically both for $L_t=\beta$ and $L_t<\beta$. 

For two-dimensional pure Yang-Mills, it is already known that our claims can be demonstrated analytically, as we will see in Sec.~\ref{sec:2d-pure-YM} where we rephrase some known facts about this case differently. For three-dimensional SU(2) pure Yang-Mills however, an analytical assessment of our claim is not feasible and hence numerical evidence is important to verify it, which we provide in the present work. 

This paper is organized as follows. In Sec.~\ref{sec:review}, we review how linear confinement with Casimir scaling follows from the random-walk property of the Wilson and Polyakov loops. For 2d pure Yang-Mills theory, this random-walk nature can easily be seen, as reviewed in Sec.~\ref{sec:2d-pure-YM}. In Sec.~\ref{sec:3dSU2}, we study 3d SU(2) pure Yang-Mills theory numerically and provide evidence for the random walk. In Appendix~\ref{sec:Hanada-Watanabe-review}, the origin of the random walk of the Polyakov loop is reviewed. 

\begin{figure}[htbp]
\begin{center}
\scalebox{0.25}{
\includegraphics{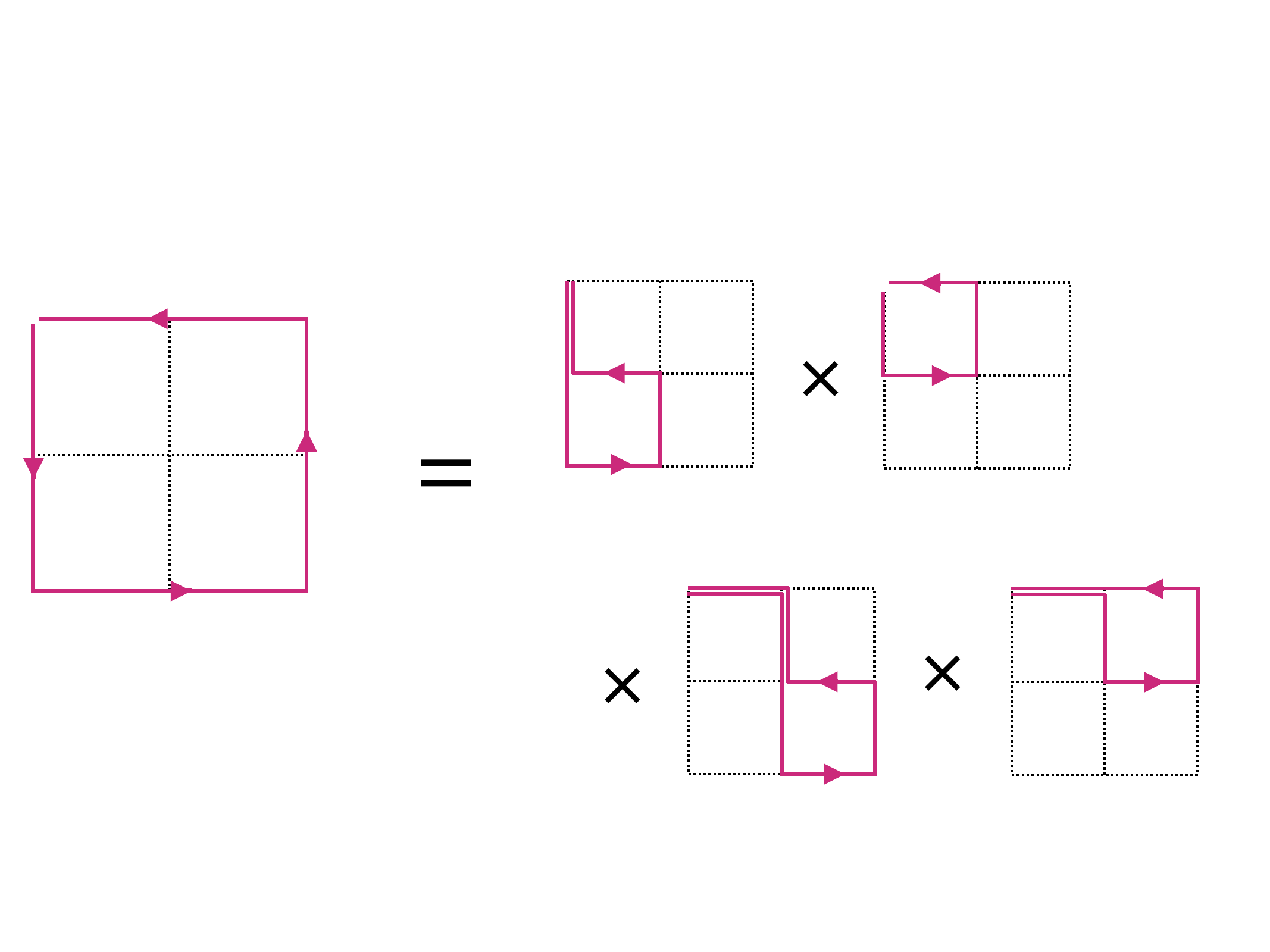}}
\end{center}
\caption{
Wilson loop expressed in terms of a product of plaquettes with `tails'.
}\label{fig:Wilson_prod}
\end{figure}

%%%%%%%%%%%%
%%%%%%%%%%%%
\section{Random walk and Casimir scaling: A review}\label{sec:review}
\hspace{0.51cm}
%%%%%%%%%%%%
%%%%%%%%%%%%
Let us see how linear confinement with Casimir scaling follows from the random-walk nature of Polyakov and Wilson loops claimed above. 
By using a short-hand notation 
\begin{align}
W^{\rm (r)}_j
=
W^{\rm (r)}_{L_t}(\vec{x}+a(j-1)\hat{\mu},\vec{x}+aj\hat{\mu})\, 
\end{align}
the $L_t\times L_s$ loop in the representation r is 
\begin{align}
 W^{\rm (r)}_1W^{\rm (r)}_2\cdots W^{\rm (r)}_K\, , 
 \end{align}
 where $Ka=L_s$. 
Suppose that the $W^{\rm (r)}_j$s are random and not correlated. Then, 
\begin{align}
\left\langle W^{\rm (r)}_1W^{\rm (r)}_2\cdots W^{\rm (r)}_K\right\rangle
=
\left\langle W^{\rm (r)}_1\right\rangle
\left\langle W^{\rm (r)}_2\right\rangle
\cdots 
\left\langle W^{\rm (r)}_K\right\rangle
=
\left(
\left\langle W^{\rm (r)}\right\rangle
\right)^K\, .
\end{align}
If, as we claimed above, $W=e^{i \sqrt{L_t} \epsilon X}$ where $X=x^\alpha T_\alpha$ is random Gaussian with variance 1 and $\epsilon$ is a small number, then
\begin{align}
\left\langle W^{\rm (r)}\right\rangle
&=
\left\langle \textbf{1} + i \sqrt{L_t} \epsilon x^\alpha T^{\rm (r)}_\alpha - \frac{L_t \epsilon^2 x^\alpha x^\beta}{2}T^{\rm (r)}_\alpha T^{\rm (r)}_\beta+\cdots\right\rangle
\nonumber\\
&=
 \textbf{1} - \frac{L_t \epsilon^2}{2}(T^{\rm (r)}_\alpha)^2+\cdots
 \nonumber\\
&=
 \textbf{1} - 2 L_t \epsilon^2 C_{\rm r} \textbf{1}+\cdots\, . 
\end{align}
Hence, for small $\epsilon$,
\begin{align}
\left\langle W^{\rm (r)}\right\rangle
\simeq
e^{-2 L_t \epsilon^2C_{\rm r}} \textbf{1} 
\end{align}
and, consequently,
\begin{align}
\left\langle W^{\rm (r)}_1\right\rangle
\left\langle W^{\rm (r)}_2\right\rangle
\cdots 
\left\langle W^{\rm (r)}_K\right\rangle
\simeq
e^{-2 L_t \epsilon^2C_{\rm r}K} \textbf{1} \, .
\end{align}
Therefore, 
\begin{align}
\left\langle W^{\rm (r)}_1W^{\rm (r)}_2\cdots W^{\rm (r)}_K\right\rangle
\simeq
e^{-2 L_t \epsilon^2C_{\rm r}K}
\textbf{1} 
\end{align}
up to $O(\epsilon^4)$ terms in the exponent. Therefore, by identifying $\epsilon$ with physical parameters as 
\begin{align}
2\epsilon^2 
=
a\sigma_0^2\, , 
\end{align}
the string tension for representation r is $C_{\rm r}\sigma_0^2$. This proves that the mentioned Casimir scaling follows from our assumptions.

The above assumption of approximately no correlation is clearly too restrictive and does not hold in three and four dimensional gauge theories. At short distances, the correlation between the $W_j$s can not be ignored, but they become uncorrelated at sufficiently large separations. As worked out in detail for the Polyakov loops in Ref.~\cite{Bergner:2023rpw}, the argument can be adapted for this case and Casimir scaling also follows from these less restrictive assumptions. What we have to confirm is only that the correlation is small.
%%%%%%%%%%%%
%%%%%%%%%%%%
\subsection{Two-dimensional pure Yang-Mills}\label{sec:2d-pure-YM}
\hspace{0.51cm}
%%%%%%%%%%%%
%%%%%%%%%%%%
Two-dimensional pure Yang-Mills theory is analytically solvable~\cite{Gross:1980he,Wadia:2012fr}. 
This theory is simple enough such that the random-walk nature stated above can easily be confirmed~\cite{Bergner:2023rpw}.

Wilson's plaquette action is given by 
\begin{align}
    S_{\rm Wilson} = -\sum_{\vec{n}} \frac{1}{2g^2a^2} \Tr{U_{\Box,\vec{n}} + U_{\Box,\vec{n}}^\dagger}\, , 
\end{align}
where the plaquette term $U_{\Box,\vec{n}}$ is 
\begin{align}
    U_{\Box,\vec{n}} = U_{t,\vec{n}} U_{x,\vec{n}+\hat{t}} U_{t,\vec{n}+\hat{x}}^\dagger U_{x,\vec{n}}^\dagger\, . 
\end{align}
We take the spatial direction to be noncompact. (We allow the temporal direction to be compact because we want to consider finite temperature.) Then, we can choose the gauge such that $U_{x,\vec{n}}=\textbf{1}$ for all $\vec{n}$, or in other words $A_x = 0$. In this gauge, the plaquette term reduces to $U_{\Box,\vec{n}} = U_{t,\vec{n}} U^\dagger_{t,\vec{n}+\hat{x}}$. 
Redefining the variables such that $W_{\vec{n}} \equiv U_{t,\vec{n}} U^\dag_{t,\vec{n}+\hat{x}} = U_{\Box,\vec{n}}$, the partition function simplifies to\footnote{Note that gauge fixing introduces no extra factor in the path integal due to properties of the Haar measure.}
\begin{align}
    Z 
    = \prod_{\vec{n}}\left(\int\mathrm{d}W_{\Vec{n}} \; \exp{\frac{1}{2g^2a^2} \Tr{W_{\Vec{n}} + W^\dag_{\Vec{n}}}}\right)\, . 
    \label{2d-partition-fnc}
\end{align}
From this expression, we see that plaquettes $W_{\vec{n}}$ behave as independent variables.

The expectation value of a Wilson loop $\mathrm{Tr}\mathcal{W}_C$ along a closed contour $C$ can then be calculated as 
\begin{align}
\langle \mathrm{Tr}\mathcal{W}_C \rangle 
=
\mathrm{Tr}\langle \mathcal{W}_C \rangle\, , 
\end{align}
where
\begin{align}
    \langle\mathcal{W}_C \rangle = \frac{1}{Z} \int\mathrm{d}W \mathcal{W}_C e^{-S_{\text{Wilson}}}\; .
\end{align}
Here, we use $\mathcal{W}_C$ to denote the closed Wilson line (Wilson loop before taking a trace), which is an $N\times N$ matrix for SU($N$) theory. 
Any closed Wilson line $\mathcal{W}_C$ can be broken down in terms of the plaquettes contained in it, as shown in Fig.~\ref{fig:Wilson_prod}. More precisely, closed Wilson lines can be expressed as a product of plaquettes with `tails'. 
Specifically in two dimensions, each plaquette $W_{\Vec{n}}$ is independent as we can see from \eqref{2d-partition-fnc}, and hence, the plaquette with `tail' is also independent.
Therefore, the closed Wilson line in two dimensions factorizes to the product of plaquettes contained in it. For a Wilson line consisting of $n_{\rm plaq.}$ plaquettes, we obtain
\begin{align}
    \langle \mathcal{W}_C\rangle 
    = 
    \left(w\cdot\textbf{1}\right)^{n_{\rm plaq.}}
    = 
    w^{n_{\rm plaq.}}\cdot\textbf{1}\, , 
\end{align}
where $w$ is the expectation value of single plaquette, i.e. $\langle W_{\Vec{n}} \rangle=w\cdot\textbf{1}$. 

When the lattice spacing $a$ is small, $W_{\vec{n}}$ localizes around $\textbf{1}$. Let us introduce a traceless Hermitian matrix $X_{\vec{n}}$ that corresponds to the field strength by 
\begin{align}
W_{\vec{n}}=e^{iaX_{\vec{n}}}\, . 
\end{align}
The action is Gaussian up to small corrections suppressed in the continuum limit: 
\begin{align}
    S_{\rm Wilson} 
    =
    \frac{1}{2g^2}
    \sum_{\vec{n}}\mathrm{Tr}X^2_{\vec{n}}
    +
    O(a^2)\, .  
\end{align}
As an approximation of the Haar measure $dW$, we use the flat measure $dX$. Therefore, $X_{\vec{n}}$ is Gaussian random and the variance is $g^2$. 
The thin loop of the form shown in Fig.~\ref{fig:basic-idea}, with one-lattice-unit extension along the spatial direction, is obtained by taking a product of $L_t/a$ plaquettes with tails. The sum of $L_t/a$ independent Gaussian random numbers with variance $g^2$ is a Gaussian random number with variance $g^2L_t/a$. Therefore, the thin loop takes the form of $e^{iga\sqrt{L_t/a}\tilde{X}}=e^{ig\sqrt{L_t a}\tilde{X}}$, where $\tilde{X}$ is Gaussian random with variance 1.

%%%%%%%%%%%%
%%%%%%%%%%%%
\section{Three-dimensional SU(2) pure Yang-Mills with gradient flow}\label{sec:3dSU2}
\hspace{0.51cm}
%%%%%%%%%%%%
%%%%%%%%%%%%
In this section, we study three-dimensional SU(2) pure Yang-Mills theory numerically.
In Sec.~\ref{sec:3d-setup}, we explain the simulation setup, specifically the choice of lattice action, parameters, and gradient flow.
The outcome of the simulations is explained in Sec.~\ref{sec:3dSU2-results}. We can see nontrivial evidence of the random-walk property. 

The eigenvalues of SU(2) matrices such as the Wilson line or Polyakov line can be written as $e^{\pm i\theta}$, where the plot of the phase $\theta$ can be restricted to $0\le\theta\le\pi$. 
%%%%%%%%%%%%
%%%%%%%%%%%%
\subsection{Simulation setup}\label{sec:3d-setup}
\hspace{0.51cm}
%%%%%%%%%%%%
%%%%%%%%%%%%
Wilson's plaquette action is given by\footnote{
In terms of Hermitian variables $A_\mu$, the link variables are $U_\mu=e^{iagA_\mu}$, and $S_{\rm Wilson} = \frac{1}{4}\sum_{\mu\neq\nu}\int d^3x{\rm Tr}F_{\mu\nu}^2$ up to higher order terms.
Another common convention uses $\frac{1}{g^2a}$ instead of $\frac{1}{2g^2a}$ as an overall factor, so that the continuum action becomes $\frac{1}{2}\int d^3x{\rm Tr}F_{\mu\nu}^2$ instead of $\frac{1}{4}\int d^3x{\rm Tr}F_{\mu\nu}^2$. This normalization was used in the series of papers by Teper and collaborators (e.g.\ Ref.~\cite{Teper:1998te}) that were used to check the validity of our simulation code. \eqref{lattice_coupling} does not change with these conventions, while \eqref{lattice_coupling_2} becomes $b=\frac{2N}{g^2a}$.  
Note also that in much of the lattice literature, including Ref.~\cite{Teper:1998te}, the letter $\beta$ for the inverse coupling. In this paper, we use $b$ for the inverse lattice coupling and $\beta$ for the inverse temperature. 
}   
\begin{align}
    S_{\rm Wilson} = -b\cdot\frac{1}{2N}\sum_{\vec{n}}\sum_{\mu\neq\nu}\mathrm{Tr}
    \left(U_{\mu,\vec{n}} U_{\nu,\vec{n}+\hat{\mu}} U_{\mu,\vec{n}+\hat{\nu}}^\dagger U_{\nu,\vec{n}}^\dagger\right)\, , 
    \label{lattice_coupling}
\end{align}
where `lattice inverse coupling' $b$ is defined as
\begin{align}
b=\frac{N}{g^2a}=\frac{N^2}{\lambda a}\, ,
    \label{lattice_coupling_2}
\end{align}
%\begin{align}
%    S_{\rm Wilson} = -\frac{1}{2g^2a}\sum_{\vec{n}}\sum_{\mu\neq\nu}\mathrm{Tr}
%    \left(U_{\mu,\vec{n}} U_{\nu,\vec{n}+\hat{\mu}} U_{\mu,\vec{n}+\hat{\nu}}^\dagger U_{\nu,\vec{n}}^\dagger\right)\, , 
%\end{align}
and $a$ is lattice spacing. The coupling constant $g^2$ has dimension of mass. 

We set the 't Hooft coupling $\lambda=g^2N$ to be 1 to define the scale of our simulations. Consequently, we measure dimensionful quantities in the unit of the 't Hooft coupling, e.g., lattice spacing $a$ and temperature $T$ are actually the dimensionless combinations $\lambda a$ and $\lambda^{-1}T$, respectively. Because we focus on $N=2$ in this paper, we have $b=\frac{4}{\lambda a}=\frac{4}{a}$.  
%%%%%%%%%%%%
%%%%%%%%%%%%
%\subsubsection*{Renormalization and smearing}
%\hspace{0.51cm}
%%%%%%%%%%%%
%%%%%%%%%%%%

When we take the continuum limit ($a\to 0$) in a three or four-dimensional theory, we need to take renormalization effects into account~\cite{Hanada:2023rlk,Bergner:2023rpw}. 
Slowly varying Haar randomness is expected after renormalization or smearing~\cite{Bergner:2023rpw}.  
In this work, we adopt the gradient flow~\cite{Luscher:2010iy,Narayanan:2006rf} to obtain a smeared gauge field and use it to calculate Polyakov and Wilson loops. We call such loops obtained by smeared field \textit{smeared loops}.
We apply gradient flow with a standard Wilson gauge action to obtain renormalized Wilson and Polyakov loops~\cite{Luscher:2010iy,Datta:2015bzm,Petreczky:2015yta}. The Wilson flow leads to renormalized quantities and this property has been used for Wilson loops and Polyakov loops before, see for example Ref.~\cite{Datta:2015bzm,Petreczky:2015yta}.
We use a fixed smearing radius in units of the gauge coupling and scale the flow time $\tau$ accordingly such that $\sqrt{8\tau}b$ is fixed.\footnote{In Refs.~\cite{Datta:2015bzm,Petreczky:2015yta}, $\tau$ and $t$ are used for imaginary time and flow time, respectively. This is the opposite of our notation.}
Our reference scale is $\tau=0.5$ at $a=1.0$, $b=4.0$. Note that the smearing does not affect the long-distance behavior and hence the string tension. 

One might be worried that the short-distance behavior of the smeared Wilson line contains too much lattice artifact. This is not a problem for the consequence of randomness we are interested in, namely the Casimir scaling of string tension is not affected by the smearing.
%%%%%%%%%%%%
%%%%%%%%%%%%
\subsection{Simulation results}\label{sec:3dSU2-results}
\hspace{0.51cm}
%%%%%%%%%%%%
%%%%%%%%%%%%
We study $\beta=T^{-1}=24$, which is in the confined phase as confirmed in Sec.~\ref{sec:confinement}. For the simulations, we use an isotropic lattice of size $n_{\rm lat}^3$  (see Table~\ref{table:simulation_parameters} for a summary of simulation parameters). Here, $\beta$ is lattice spacing $a$ times the number of lattice sites along the temporal direction $n_{\rm{lat}}$ i.e., $\beta=an_{\rm{lat}}$. 

We take the size of the Wilson loop $L_t\times L_s$, where $L_t=an_t$ and $L_s=an_s$. Several different values of lattice spacing are used. 
As we will see in Sec.~\ref{sec:confinement}, the tadpole-improved lattice spacing~\cite{Lepage:1996jw,Teper:1998te} $a_{\rm I}$ enables us to study the continuum limit more efficiently. Therefore, we use $a_{\rm I}$ instead of $a$ for the investigation of the random-walk property in Sec.~\ref{sec:3dSU2-RandomWalk}.  
The improved lattice spacing is defined as $a_{\rm I}\equiv \frac{aN}{\langle\textrm{plaquette}\rangle}=\frac{4N}{b\langle\textrm{plaquette}\rangle}$, where the average plaquette $\langle\textrm{plaquette}\rangle$ is obtained at the same simulation parameters. (In the continuum limit, the ratio between the improved lattice spacing $a_{\rm I}$ and the original lattice spacing $a$ becomes 1.)

\begin{table}
 \centering
  \begin{tabular}{|c||c|c|c|c|}
        \hline
        $\beta=T^{-1}$ & $a$ & $b=\frac{4}{a}$ & lattice size\\
        \hline
        \hline
         24 & 2.000 & 2 & $12^3$\\
         24 & 1.333 & 3 & $18^3$\\
         24 & 1.000 & 4 & $24^3$\\
         24 & 0.800 & 5 & $30^3$\\
         24 & 0.666 & 6 & $36^3$\\
         24 & 0.571 & 7 & $42^3$\\
         24 & 0.500 & 8 & $48^3$\\
        \hline
  \end{tabular}
  \caption{Simulation parameters }\label{table:simulation_parameters}
\end{table}

%%%%%%%%%%%%
%%%%%%%%%%%%
\subsubsection{Confined behavior of Polyakov loops}\label{sec:confinement}
\begin{figure}[htbp]
\begin{center}
\scalebox{0.45}{
\includegraphics{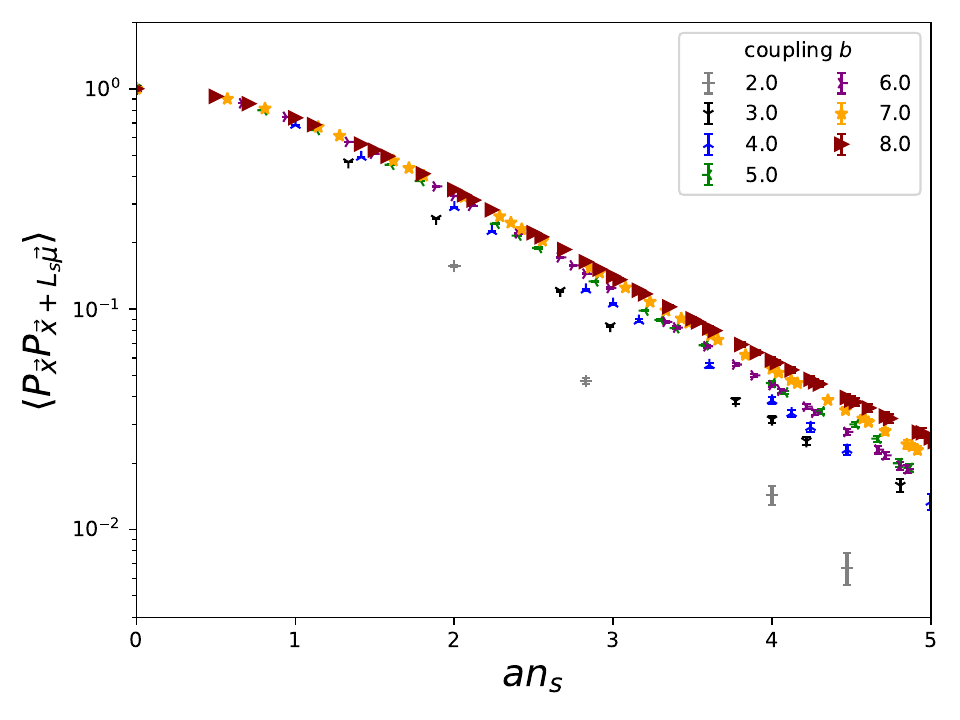}}
\scalebox{0.45}{
\includegraphics{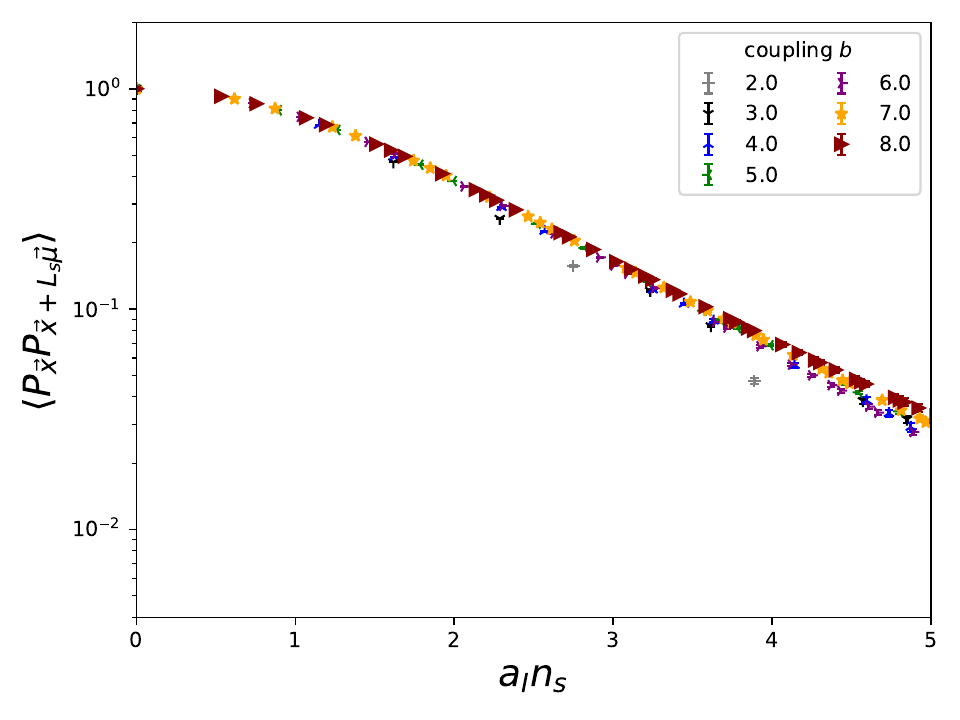}}
\end{center}
\caption{
Logarithmic plot of Polyakov loop correlator at distance $L_s$.
In the right panel, $a_{\rm I}=a\times\frac{N}{\langle\textrm{plaquette}\rangle}$ was used to determine $L_s$. 
}\label{fig:Polyakov_loop_continuum_limit}
\end{figure}

\hspace{0.51cm}
%%%%%%%%%%%%
%%%%%%%%%%%%
Fig.~\ref{fig:Polyakov_loop_continuum_limit} shows the two-point function of smeared Polyakov loops with spatial separation $L_s$ at $\beta=T^{-1}=24$.
In the right panel, the tadpole-improved lattice spacing $a_{\rm I}$ was used. With this improvement, there is a faster convergence to the continuum limit as $b\to\infty$. Therefore, we will use the improved lattice spacing later in Sec.~\ref{sec:3dSU2-RandomWalk}. We can see a clear exponential decay at large $L_s$, which shows that $\beta=24$ is in the confined phase. 

\begin{figure}[htbp]
\begin{center}
\scalebox{0.6}{
\includegraphics{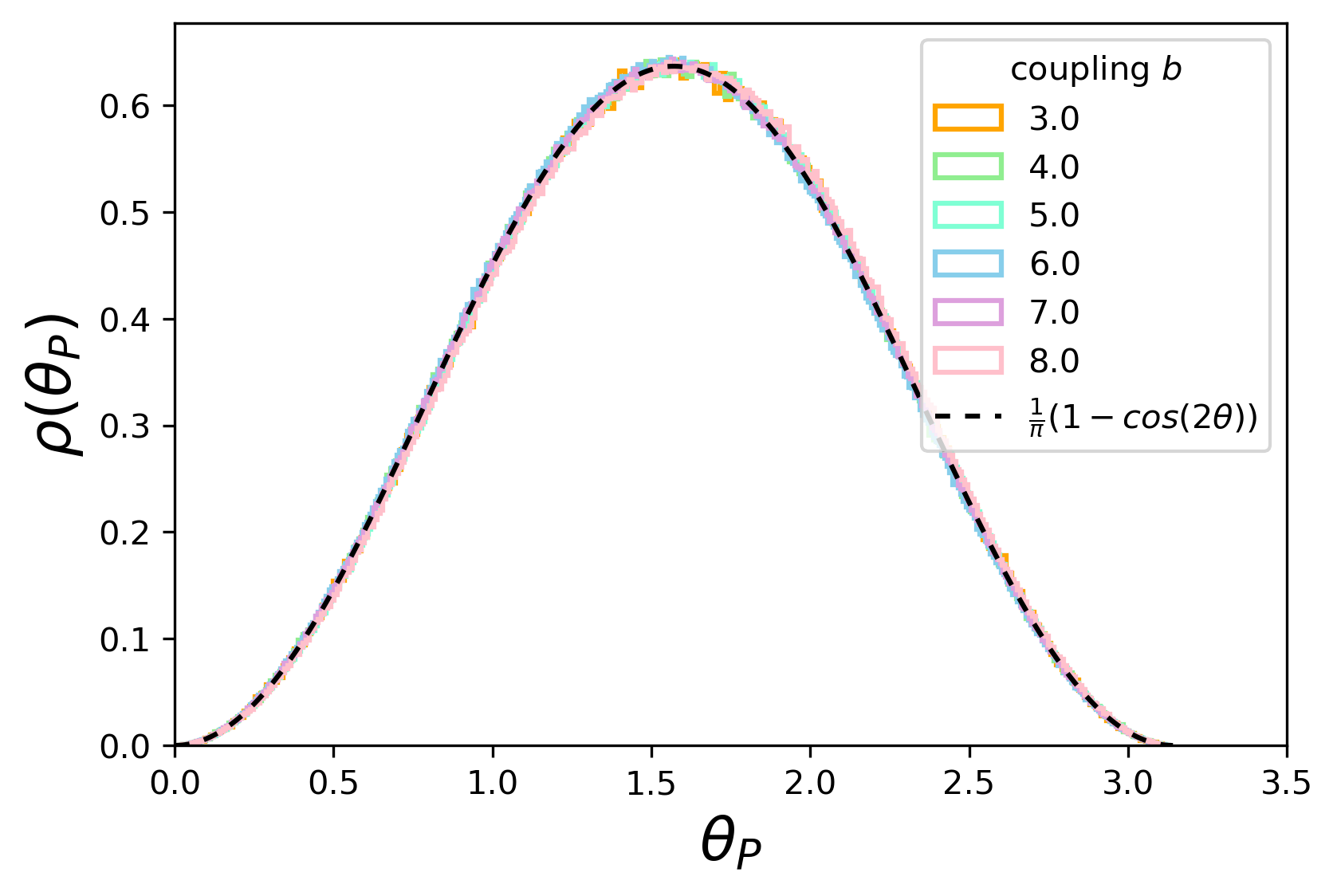}}
\end{center}
\caption{
Distribution of phase of Polyakov loop at various couplings b, all at the same inverse temperature $\beta = T^{-1} = 24$. The distribution can clearly be seen to be consistent with the Haar random distribution $\rho(\theta) = \frac{1}{\pi}(1-\cos{2\theta})$, also plotted here as a dashed line. 
}\label{fig:Polyakov-phase}
\end{figure}

Furthermore, as shown in Fig.~\ref{fig:Polyakov-phase}, the distribution of the phase of the smeared Polyakov loop coincides with the Haar-random distribution. This provides further evidence that $\beta=24$ is in the confined phase~\cite{Hanada:2023krw,Hanada:2023rlk}.

%%%%%%%%%%%%
%%%%%%%%%%%%
\subsubsection{Random walk}\label{sec:3dSU2-RandomWalk}
\begin{figure}[htbp]
\begin{center}
\scalebox{0.2}{
\includegraphics{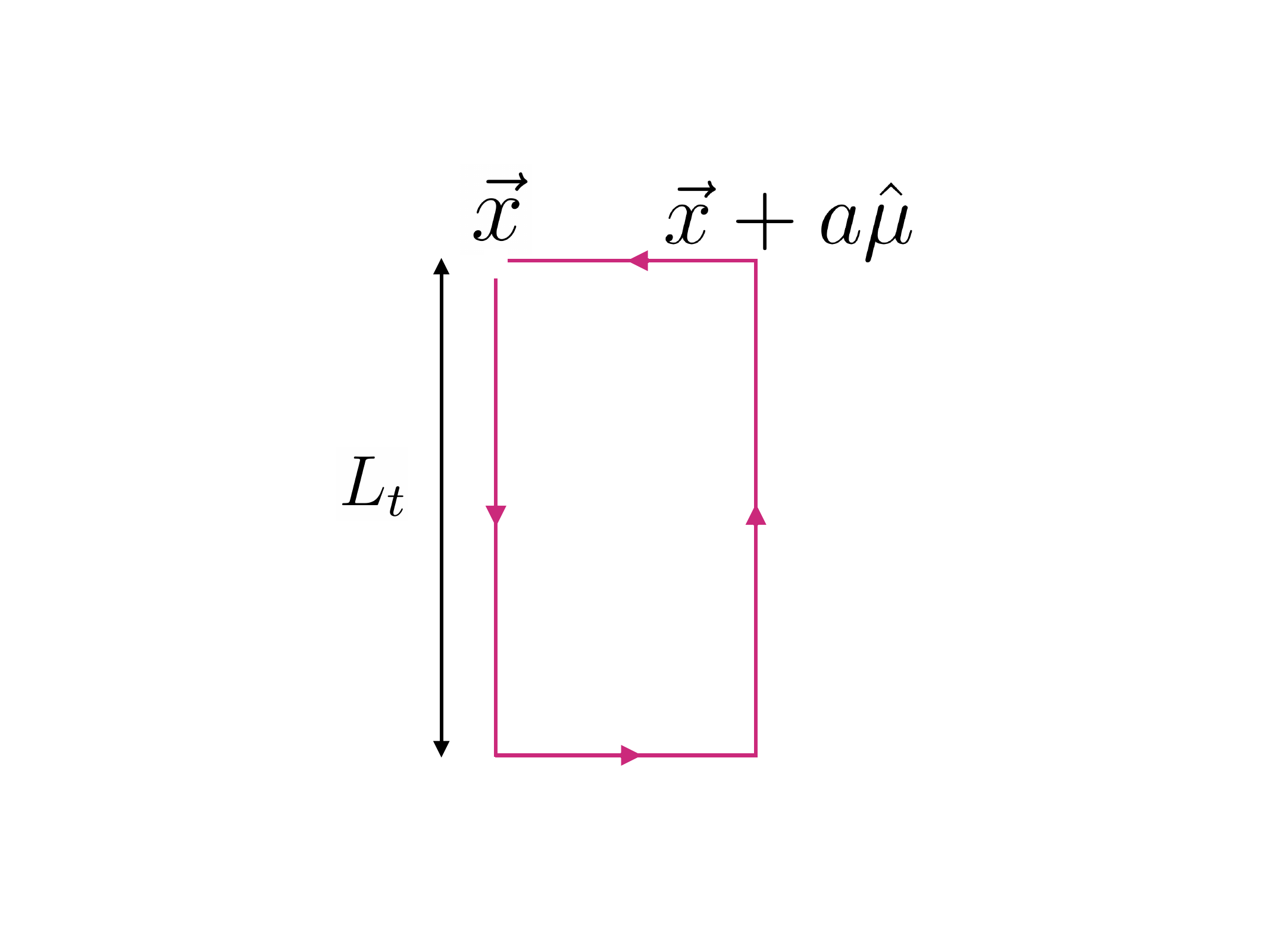}}
\scalebox{0.2}{
\includegraphics{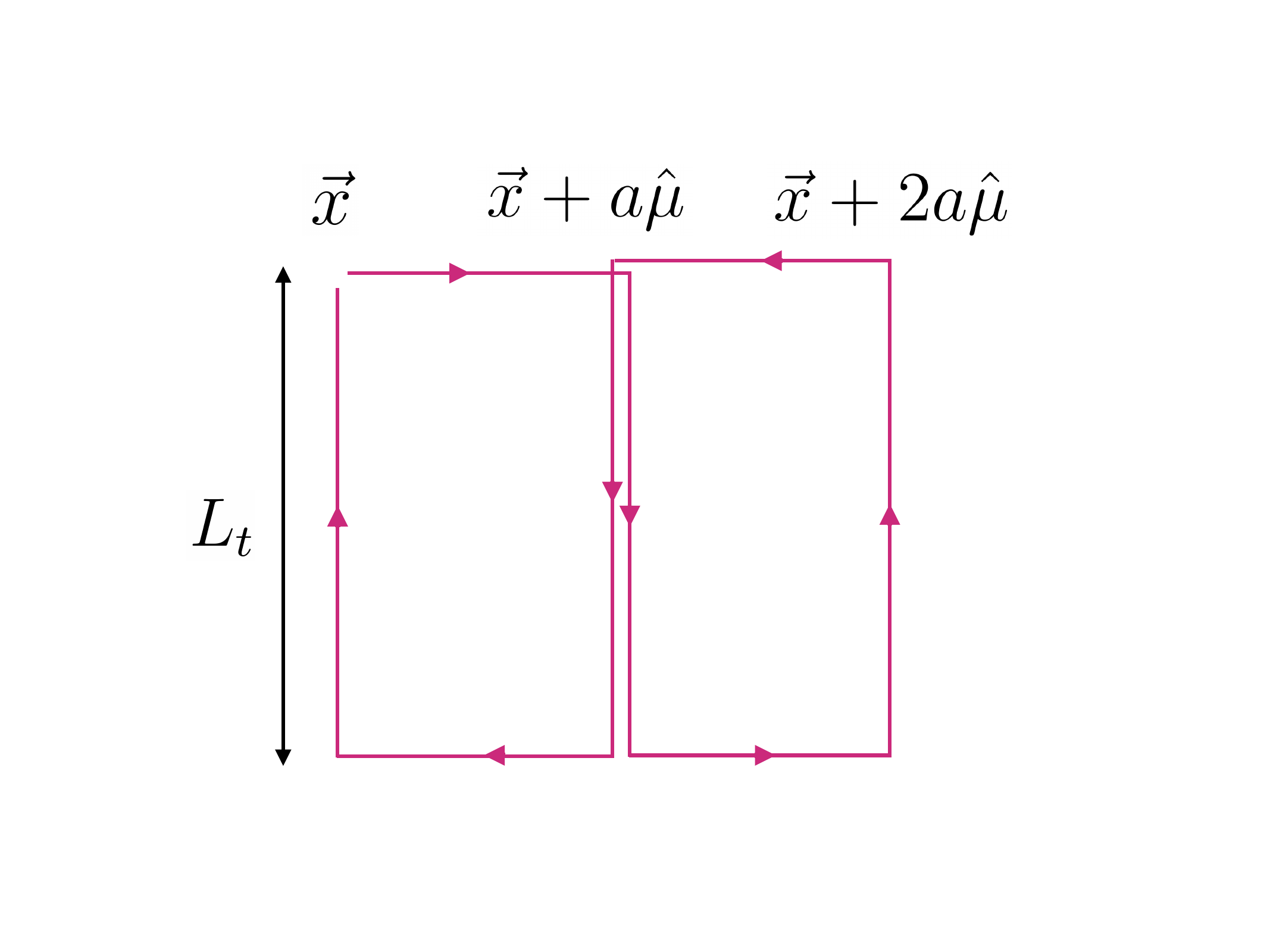}}
\end{center}
\caption{
[Left] A Wilson loop which is equivalent to $W_{L_t}(\vec{x},\vec{x}+a\hat{\mu})$ up to gauge transformation.
[Right] A Wilson loop which is equivalent to $(W_{L_t}(\vec{x},\vec{x}+a\hat{\mu}))^{-1}\cdot W_{L_t}(\vec{x}+a\hat{\mu},\vec{x}+2a\hat{\mu})$ up to gauge transformation.
For simplicity, we call them $W(L_t\times a)$ and $A(L_t\times a)$, respectively.
}\label{fig:Wilson_lattice_test}
\end{figure}

\hspace{0.51cm}
%%%%%%%%%%%%
%%%%%%%%%%%%
To provide nontrivial evidence for the main claims, we study two different classes of Wilson lines with contours as depicted in Fig.~\ref{fig:Wilson_lattice_test}:  a Wilson line which is equivalent to $W_{L_t}(\vec{x},\vec{x}+\hat{\mu})$, which we also call $W(L_t\times a)$, and a Wilson line which is equivalent to $(W_{L_t}(\vec{x},\vec{x}+a\hat{\mu}))^{-1}\cdot W_{L_t}(\vec{x}+a\hat{\mu},\vec{x}+2a\hat{\mu})$ up to gauge transformations, which we call $A(L_t,a)$. 
As discussed in Sec.~\ref{sec: Intro},  $L_s/a$ of the thin Wilson loops $W(L_t\times a)$ can be combined to obtain a Wilson loop of spatial length $L_s$. Our main claim \eqref{eq:WX} is related to the scaling of the phases $\theta_W$ and their distribution, which provides information about $X$. $A(L_t,a)$, on the other hand, measures the correlation between two adjacent thin Wilson loops. It provides information about the derivative of $X$ with respect to the position.\footnote{Note that in all cases we take values without any volume averaging.}

\begin{figure}[htbp]
\begin{center}
\scalebox{0.8}{
\includegraphics{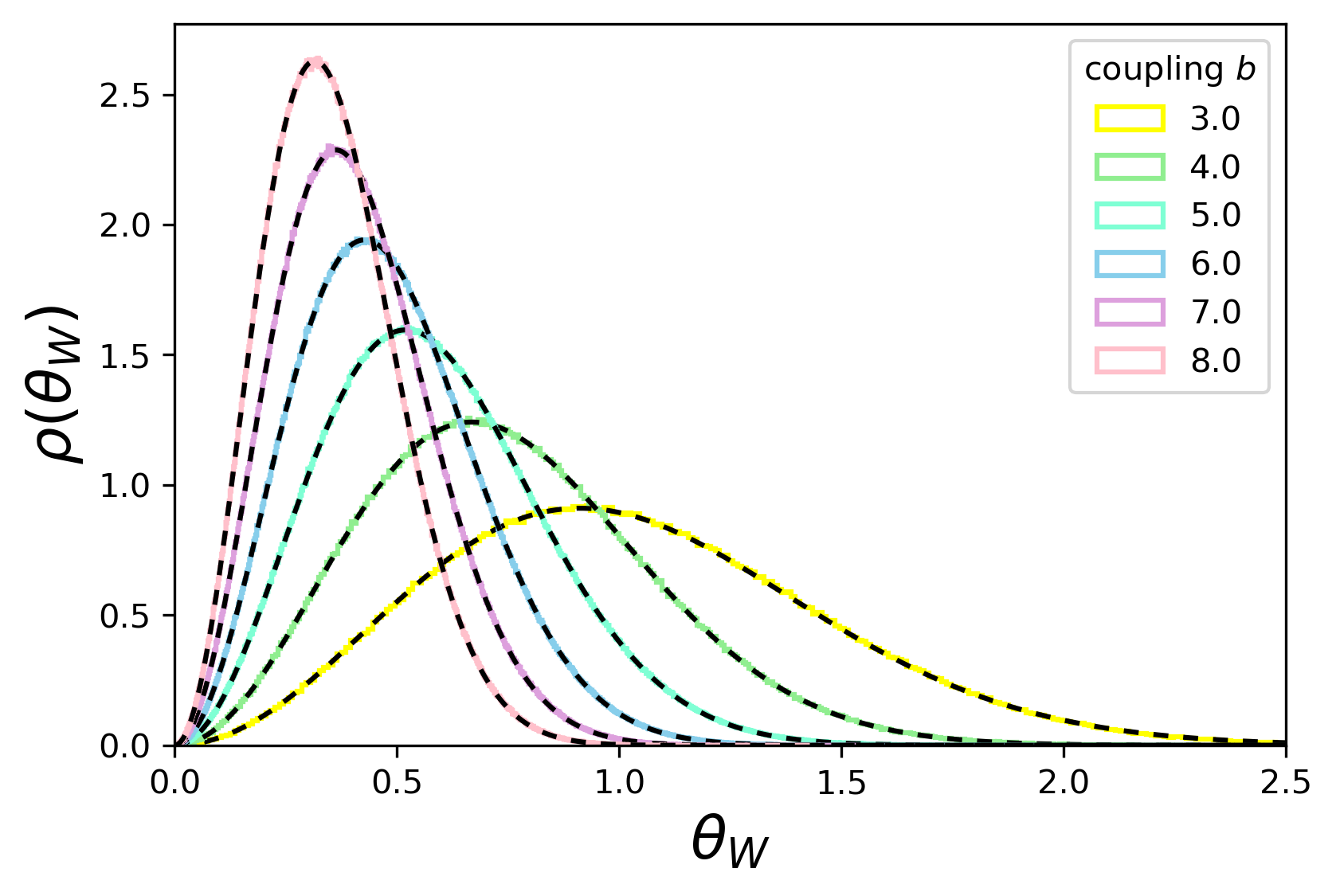}}
\scalebox{0.8}{
\includegraphics{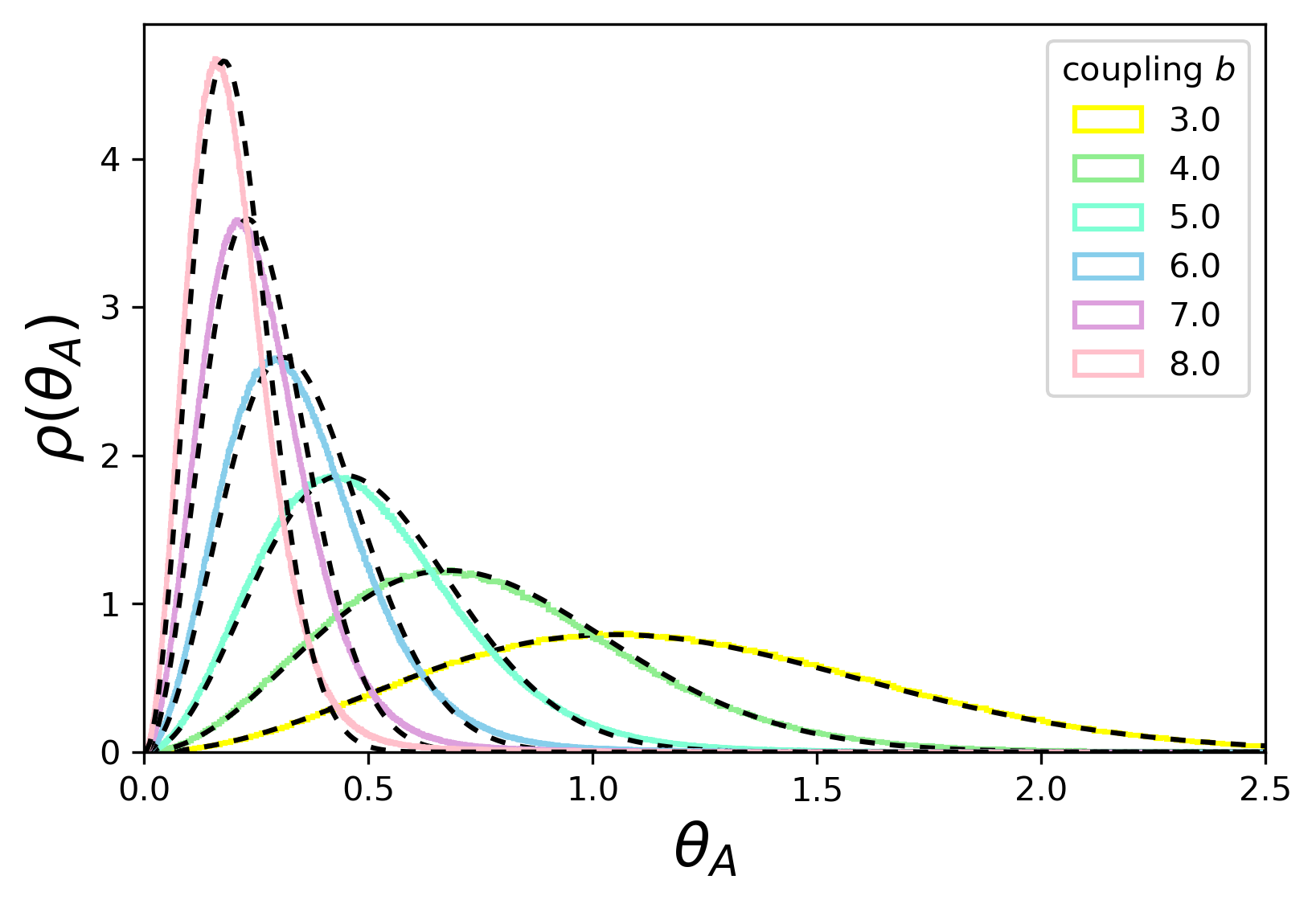}}
\end{center}
\caption{
Distribution of $\theta_W$ (top) and $\theta_A$ (bottom) obtained from $W(L_t=\beta,a)$ and $A(L_t=\beta,a)$, which are denoted by $\rho(\theta_W)$ and $\rho(\theta_A)$, are shown. The GUE Wigner surmise rescaled by a factor $\langle\theta\rangle$ (see eq.~\ref{GUEWig}) is also shown as black  dashed lines. 
$\rho(\theta_W)$ agrees well with the GUE Wigner surmise. We can see some disagreement between $\rho(\theta_A)$ and the Wigner surmise (bottom) at finer lattice spacing (larger lattice coupling $b$).
}\label{fig:W-1*W-Wigner}
\end{figure}

\begin{figure}[htbp]
\begin{center}
\scalebox{0.45}{
\includegraphics{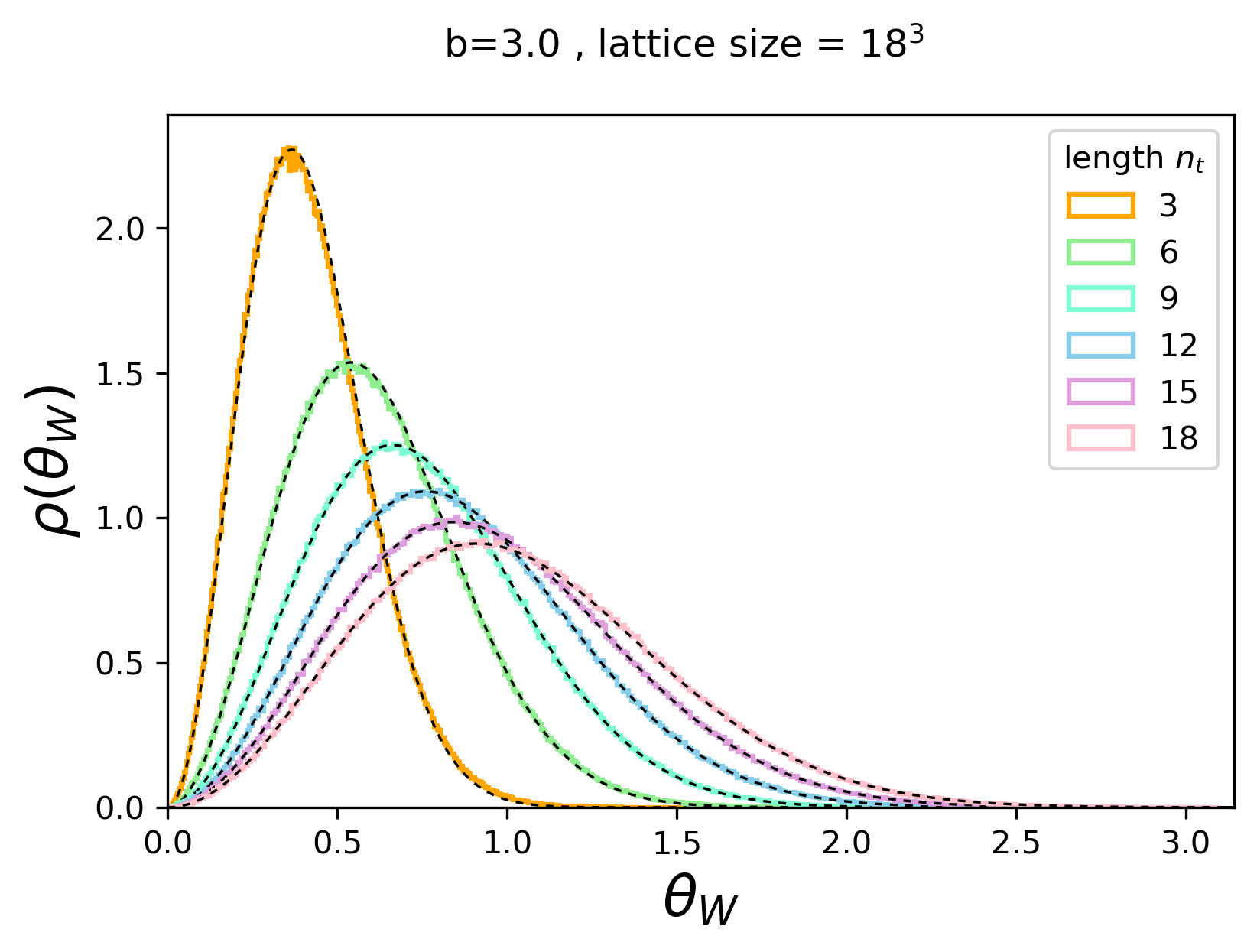}}
\scalebox{0.45}{
\includegraphics{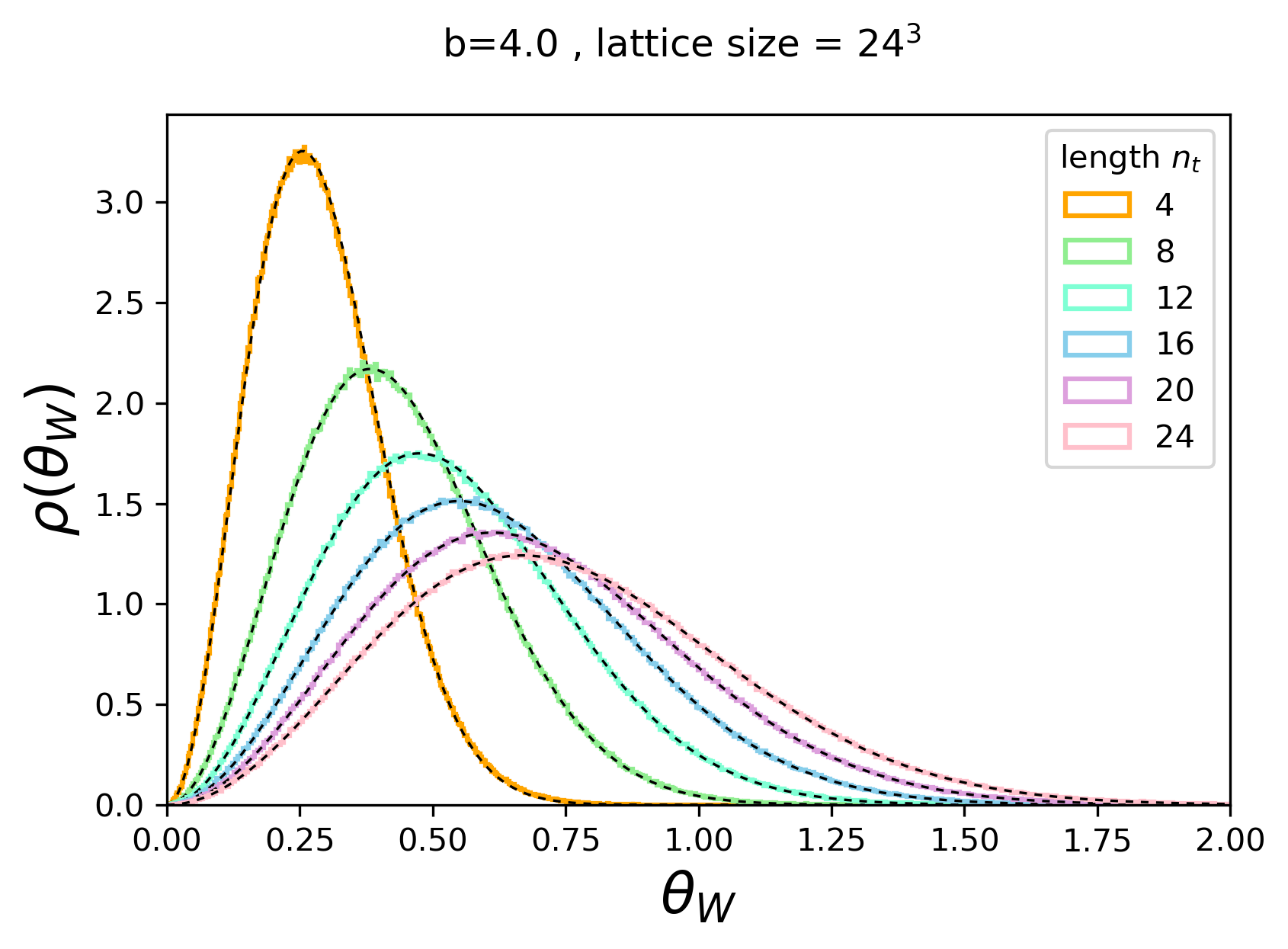}}
\scalebox{0.45}{
\includegraphics{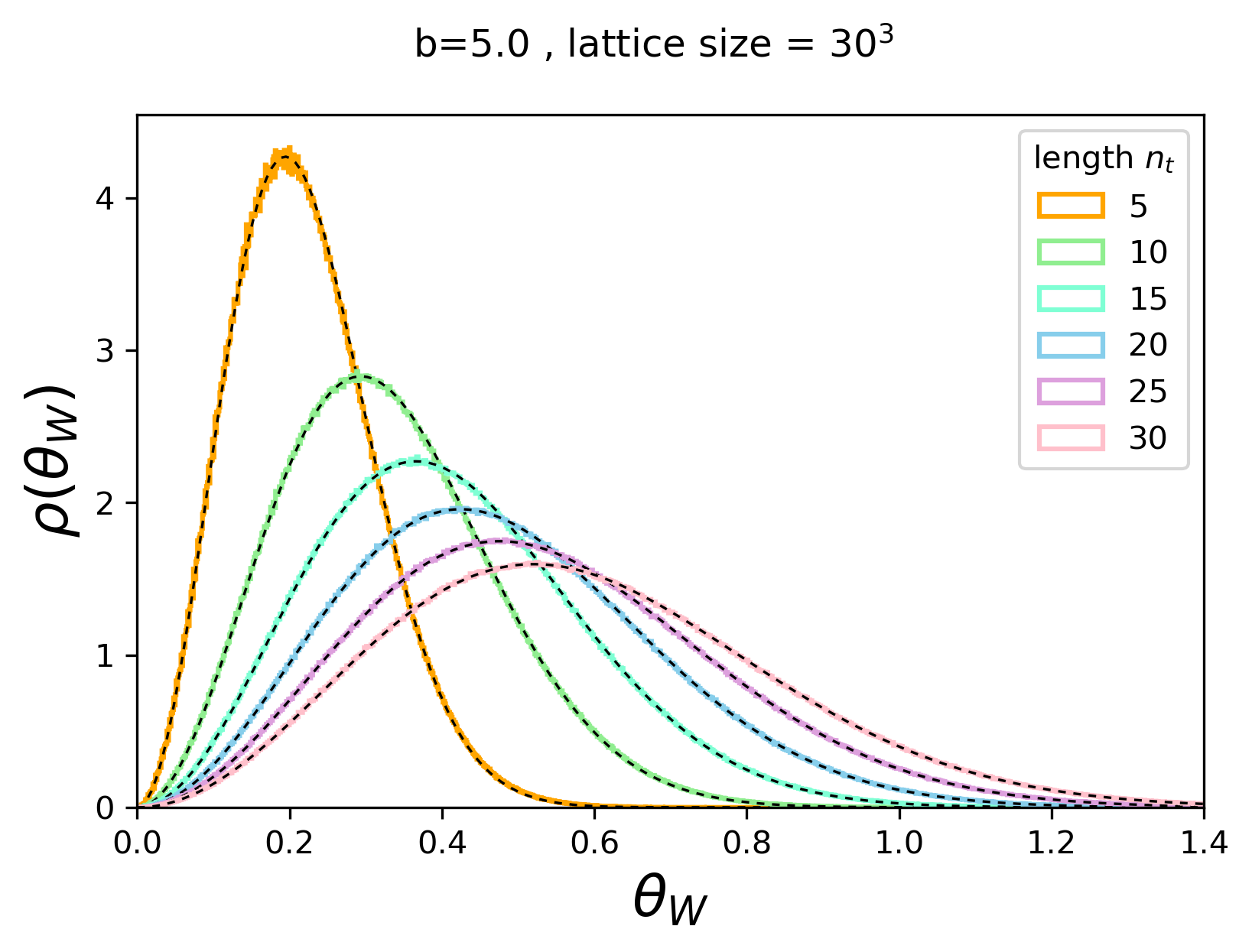}}
\scalebox{0.45}{
\includegraphics{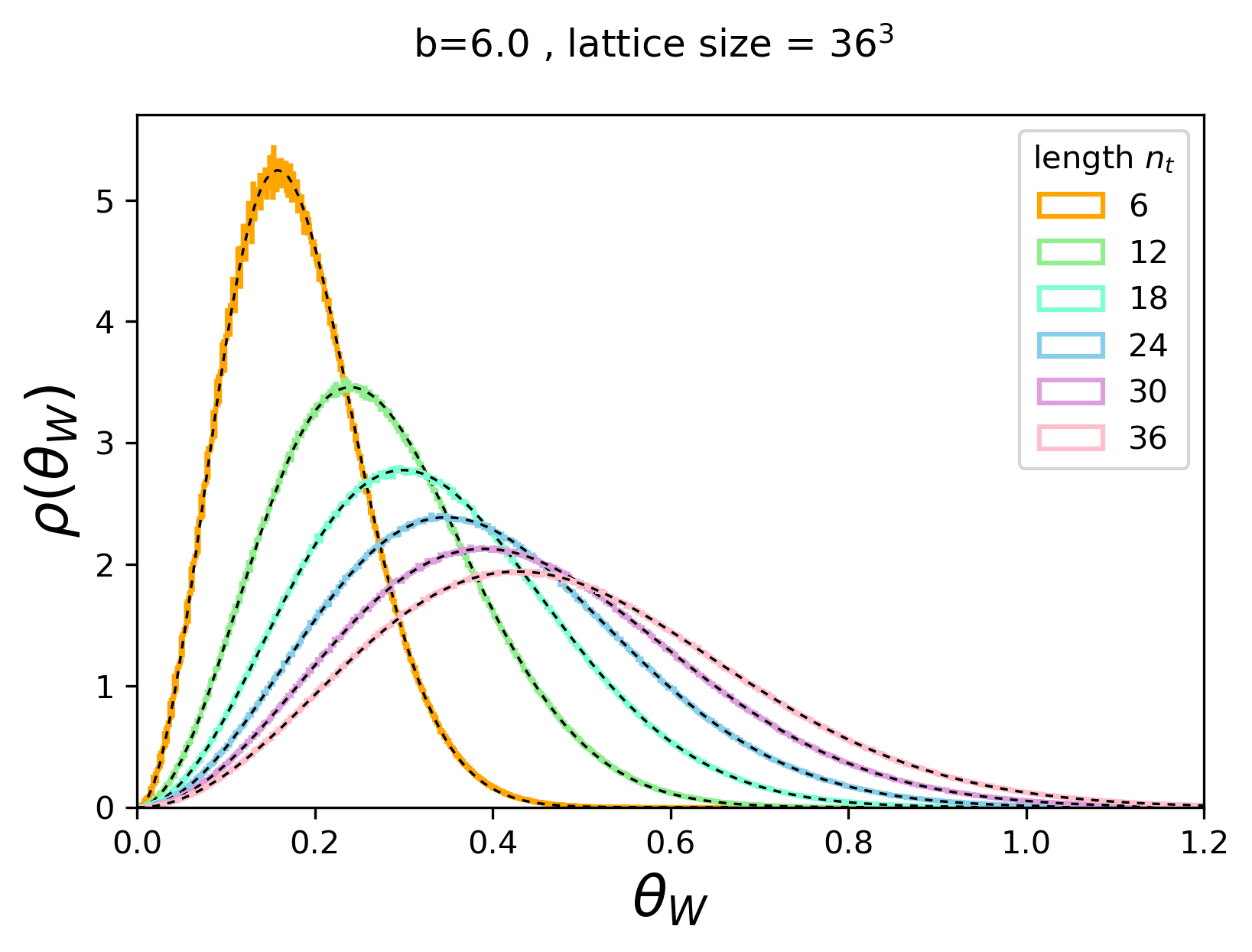}}
\scalebox{0.45}{
\includegraphics{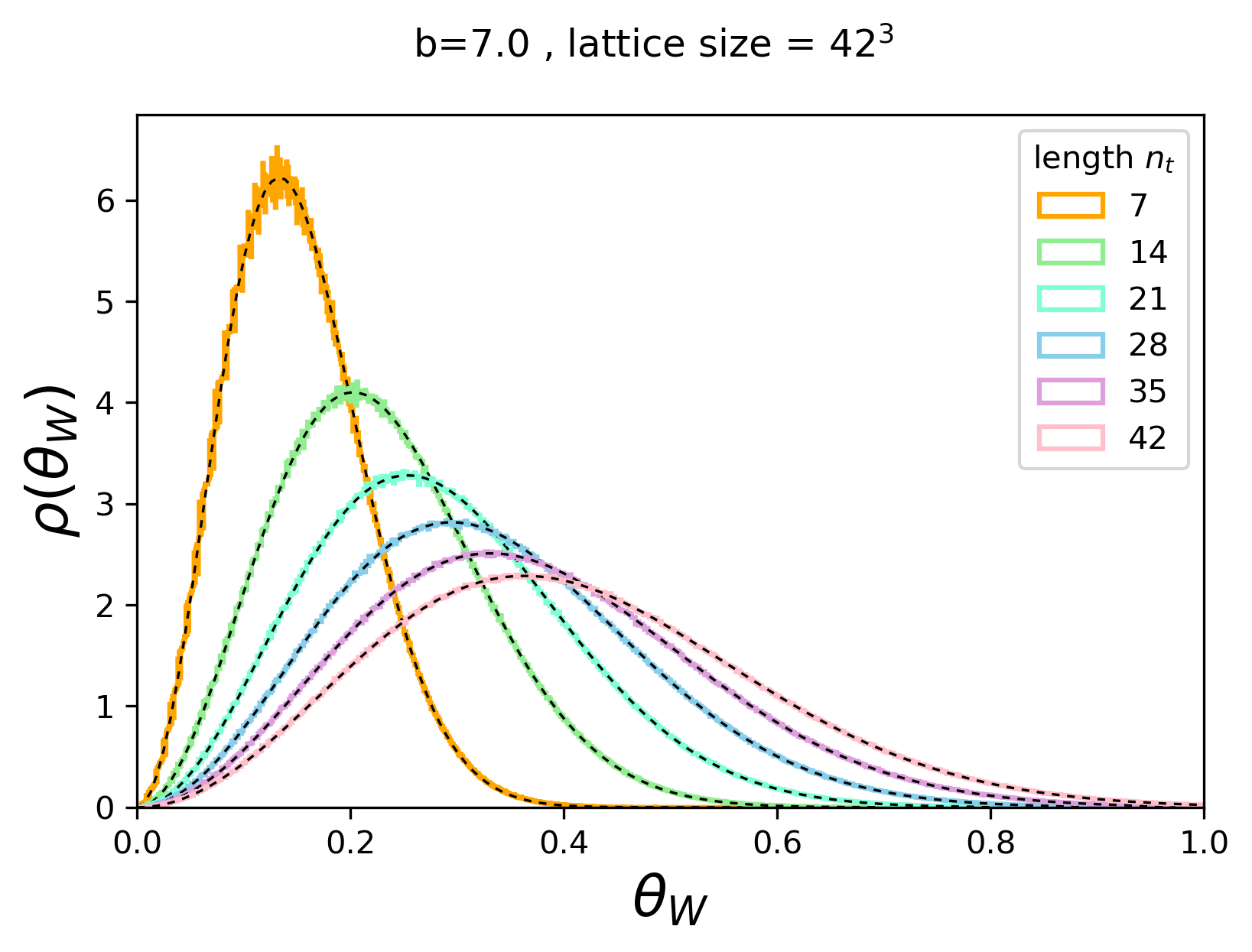}}
\scalebox{0.45}{
\includegraphics{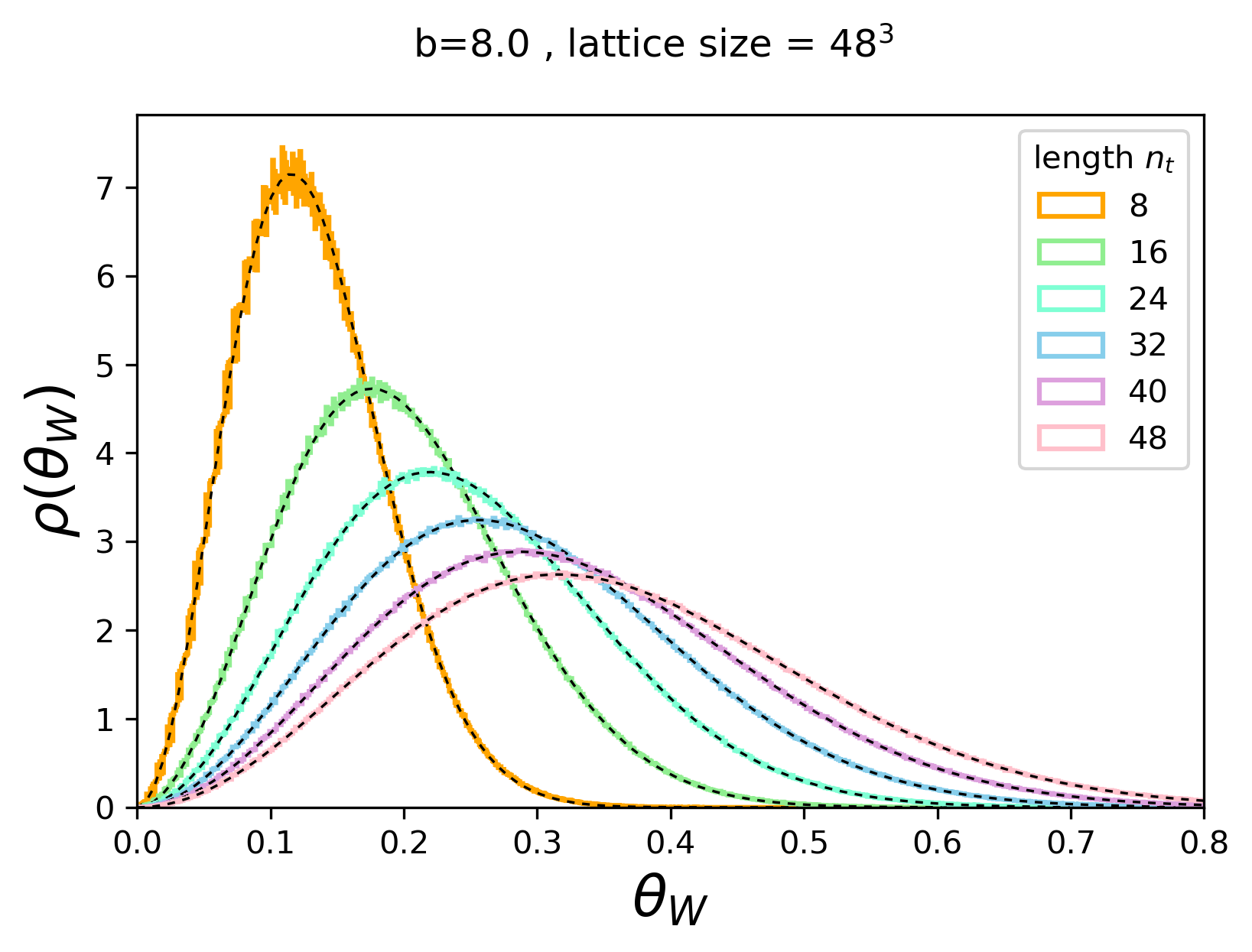}}
\end{center}
\caption{
Comparison of GUE Wigner surmise rescaled by a factor of $\langle \theta_W \rangle$ (see eq.~\ref{GUEWig}) and distribution of $\theta_W$ from $W(L_t\times a)$ , for inverse temperature $\beta = 24$ at various couplings. The phase distribution almost overlaps with the rescaled Wigner surmise (shown above as dashed lines) which supports the Gaussian-random nature of $W(L_t\times a)$. 
}\label{fig:W-Wigner}
\end{figure}

According to our conjecture, $W(L_t \times a)$  should be random matrices not just at $L_t=\beta$ but also at $L_t<\beta$. In Fig.~\ref{fig:W-1*W-Wigner}, we plotted the phases $\theta_W$ for several values of lattice spacing $a$. They follow to a very good precision the Gaussian Unitary Ensemble (GUE) Wigner surmise,
\begin{align}\label{GUEWig}
\rho_{\rm GUE}(s)
=
\frac{32s^2}{\pi^2}e^{-4s^2/\pi}\, ,
\end{align}
with the identification $s=\frac{\theta}{\langle\theta\rangle}$, and the width $\langle\theta\rangle$ decreases with the lattice spacing. This agreement with the GUE Wigner surmise indicates Gaussian randomness of the $\beta=2$ universality class~\cite{Wigner1957}. 
In Fig.~\ref{fig:W-Wigner}, we extend the analysis of the $\theta_W$ distribution to different values of $L_t$ ($L_t=4,8,12,16,20$, and $24$) at a fixed lattice spacing $a=0.5$ ($\beta=24$, on a $48\times 64^2$ lattice). In all cases, the distribution is matched with remarkable accuracy.

\begin{figure}[htbp]
\begin{center}
\scalebox{0.47}{
\includegraphics{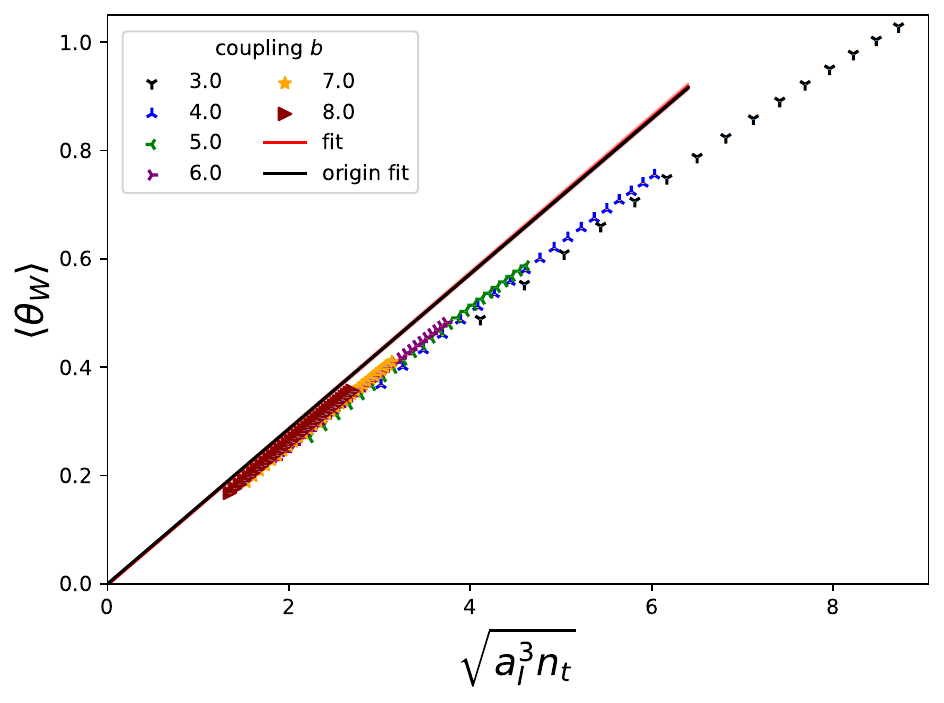}}
\scalebox{0.45}{
\includegraphics{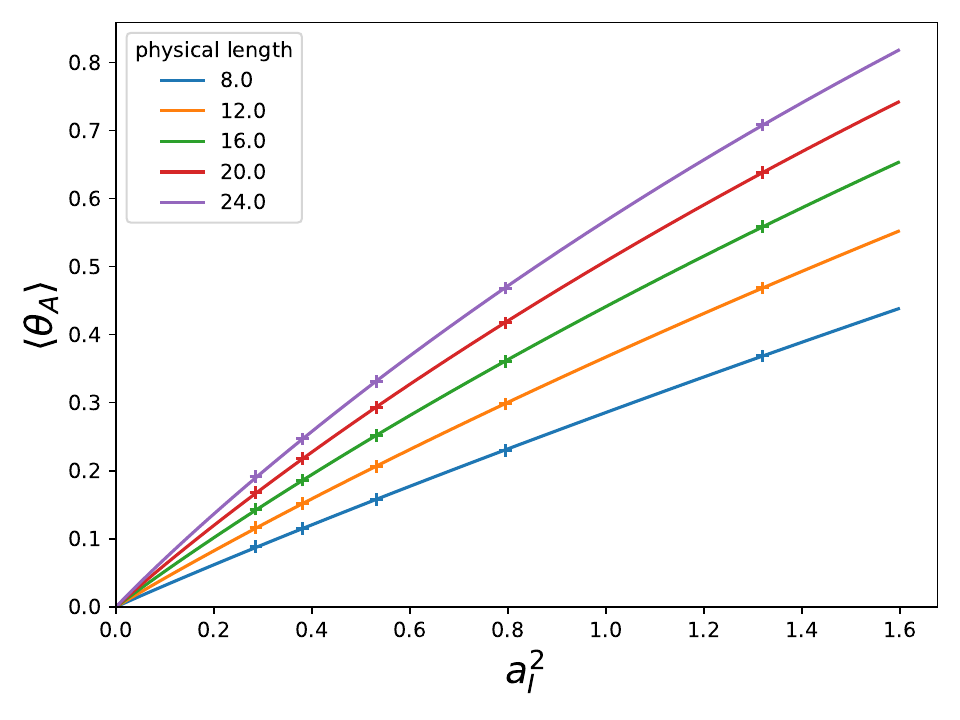}}
\scalebox{0.45}{
\includegraphics{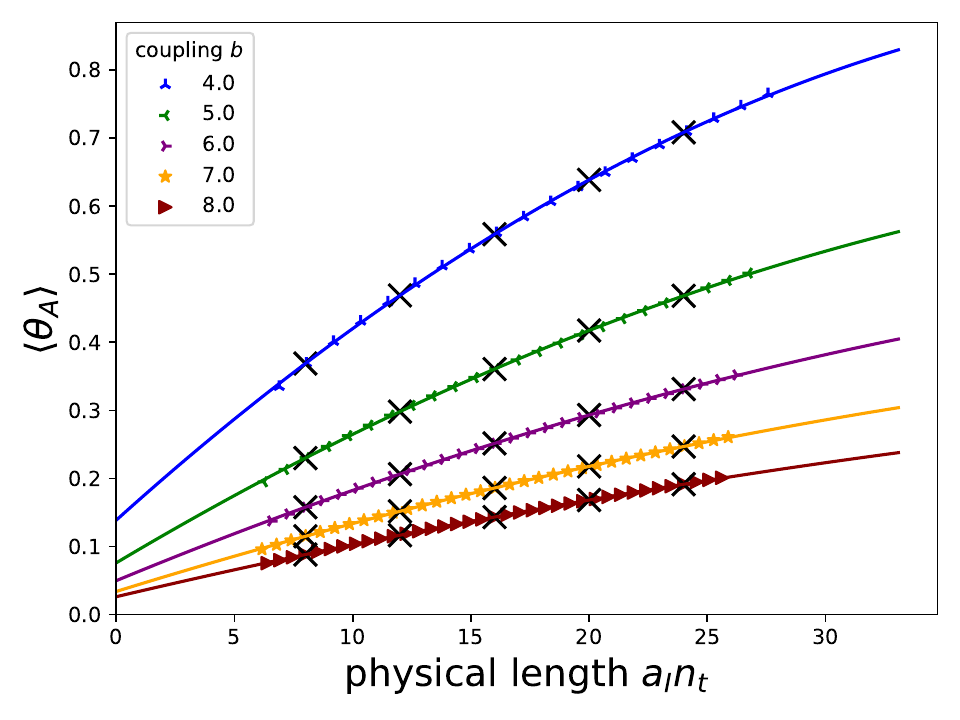}}
\end{center}
\caption{
[Top, Left]$\langle\theta_W\rangle$ vs $\sqrt{a_{\rm I}^3n_t}$. Because we are interested only in sufficiently large $L_t$, we used only loops with $a_{\rm I}n_t\ge 6$. We show the continuum fit (red), fitted against the ansatz $y=k_0 + k_v v + k_a a_I + k_{av} a_I v$. This agrees well with a similar fit forced through the origin (black), confirming physical expectations. 
[Top, Right]$\langle\theta_A\rangle$ vs $a_{\rm I}^2$, with curves fitted to the ansatz $y=k_1 a_I^2 + k_2 a_I^3$. 
[Bottom] To estimate $\langle\theta_A\rangle$ at $L_t=8,12,16,20$ and $24$, we fit $\langle\theta_A\rangle$ at each fixed value of $b$ as a function of $L_t=a_{\rm I}n_t$, using the ansatz $y=k_0 + k_1 L_t + k_2 L_t ^ 2$. Numerical uncertainties are smaller than the size of the data points.
}\label{fig:width}
\end{figure}

Relation \eqref{eq:WX} predicts that $\langle\theta_W\rangle$ scales as $a\sqrt{L_t}$ in the continuum limit. In the left panel of Fig.~\ref{fig:width}, $\langle\theta_W\rangle$ is shown as a function of $\sqrt{a_{\rm I}^3n_t}$, where only sufficiently large $L_t$ ($a_{\rm I}n_t\ge 6$) is included in the analysis. From the fit we observe $\langle\theta_W\rangle\propto\sqrt{a_{\rm I}^3n_t}$ for small lattice spacing, which confirms the expected scaling.

\begin{figure}[htbp]
\begin{center}
\scalebox{0.45}{
\includegraphics{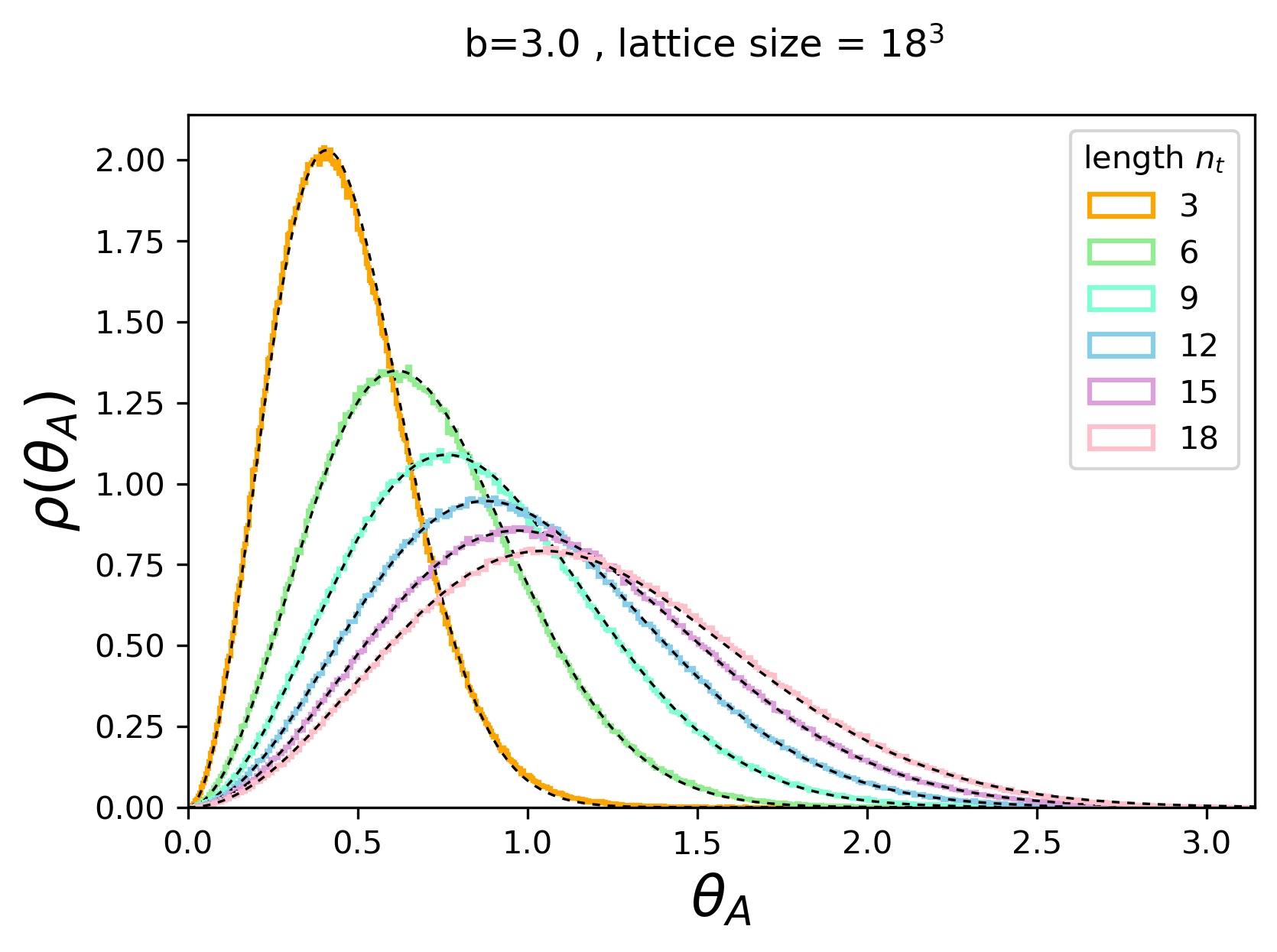}}
\scalebox{0.45}{
\includegraphics{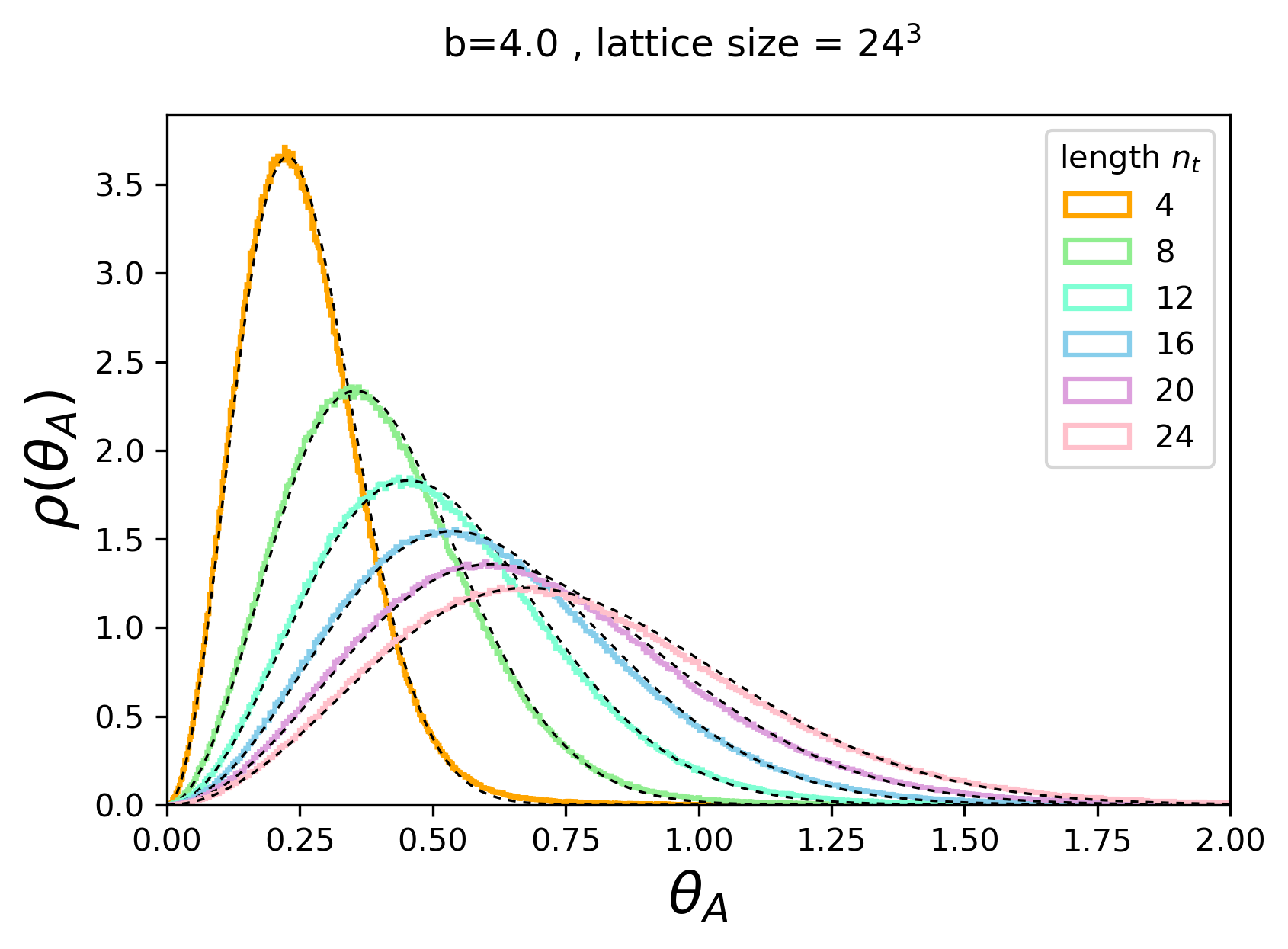}}
\scalebox{0.45}{
\includegraphics{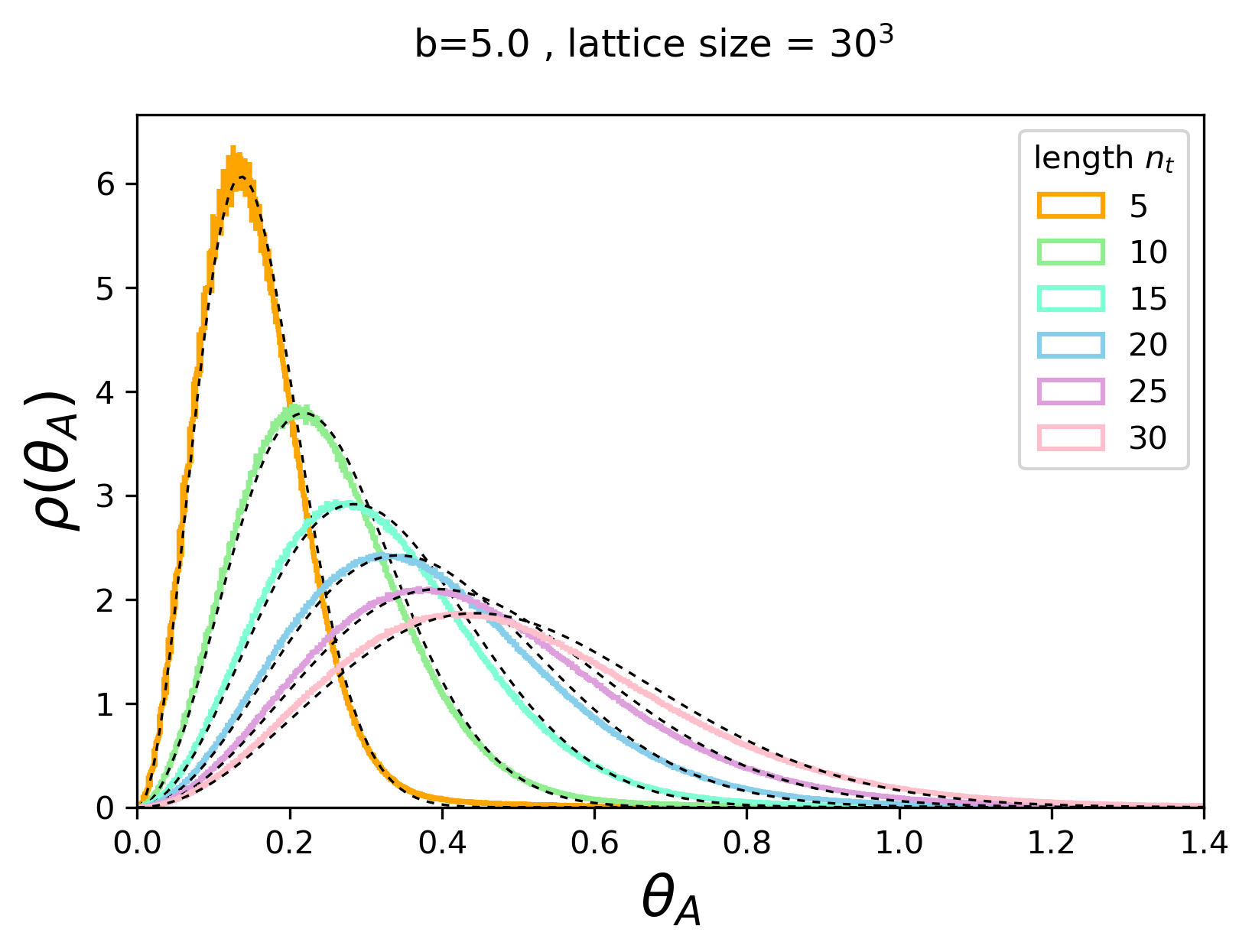}}
\scalebox{0.45}{
\includegraphics{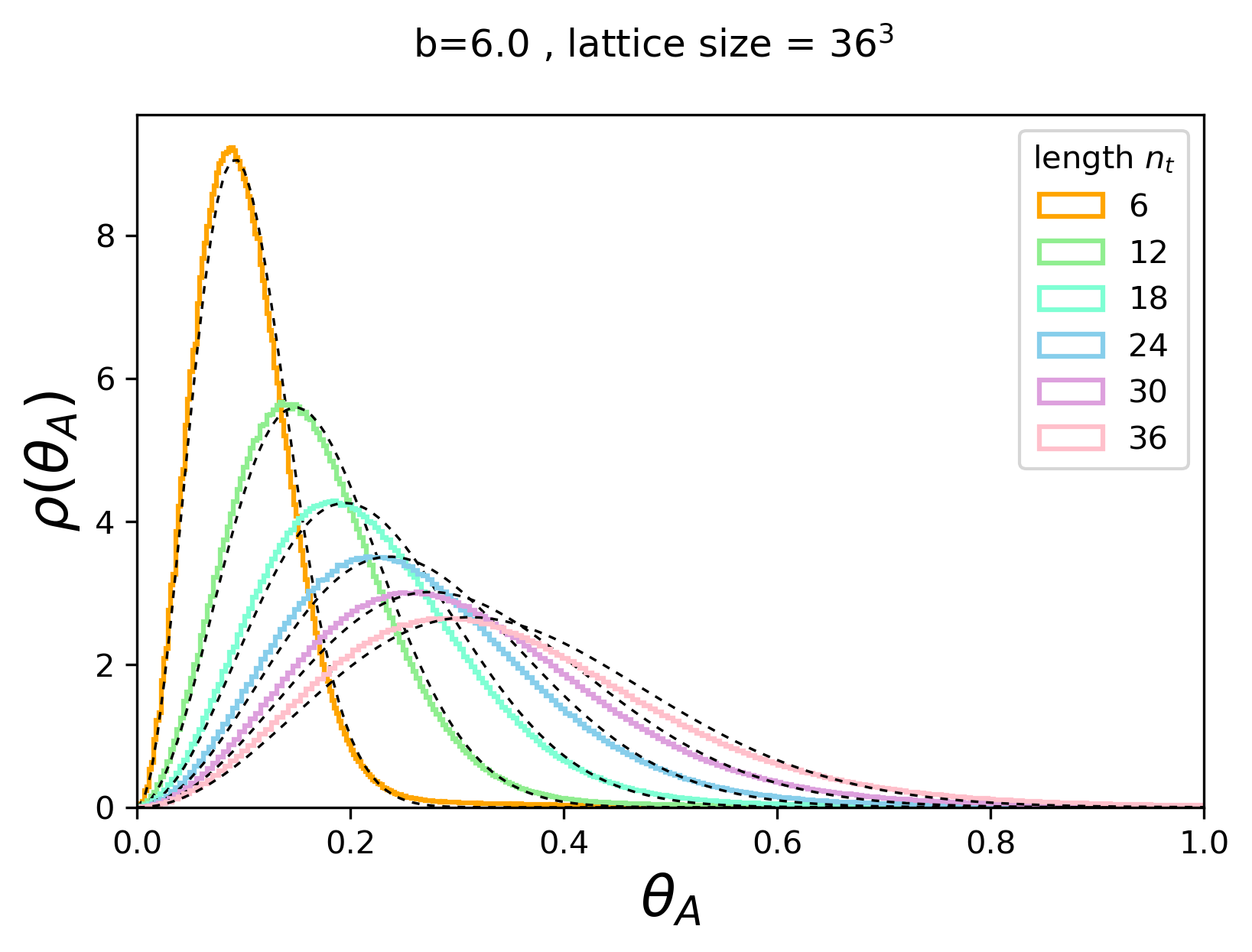}}
\scalebox{0.45}{
\includegraphics{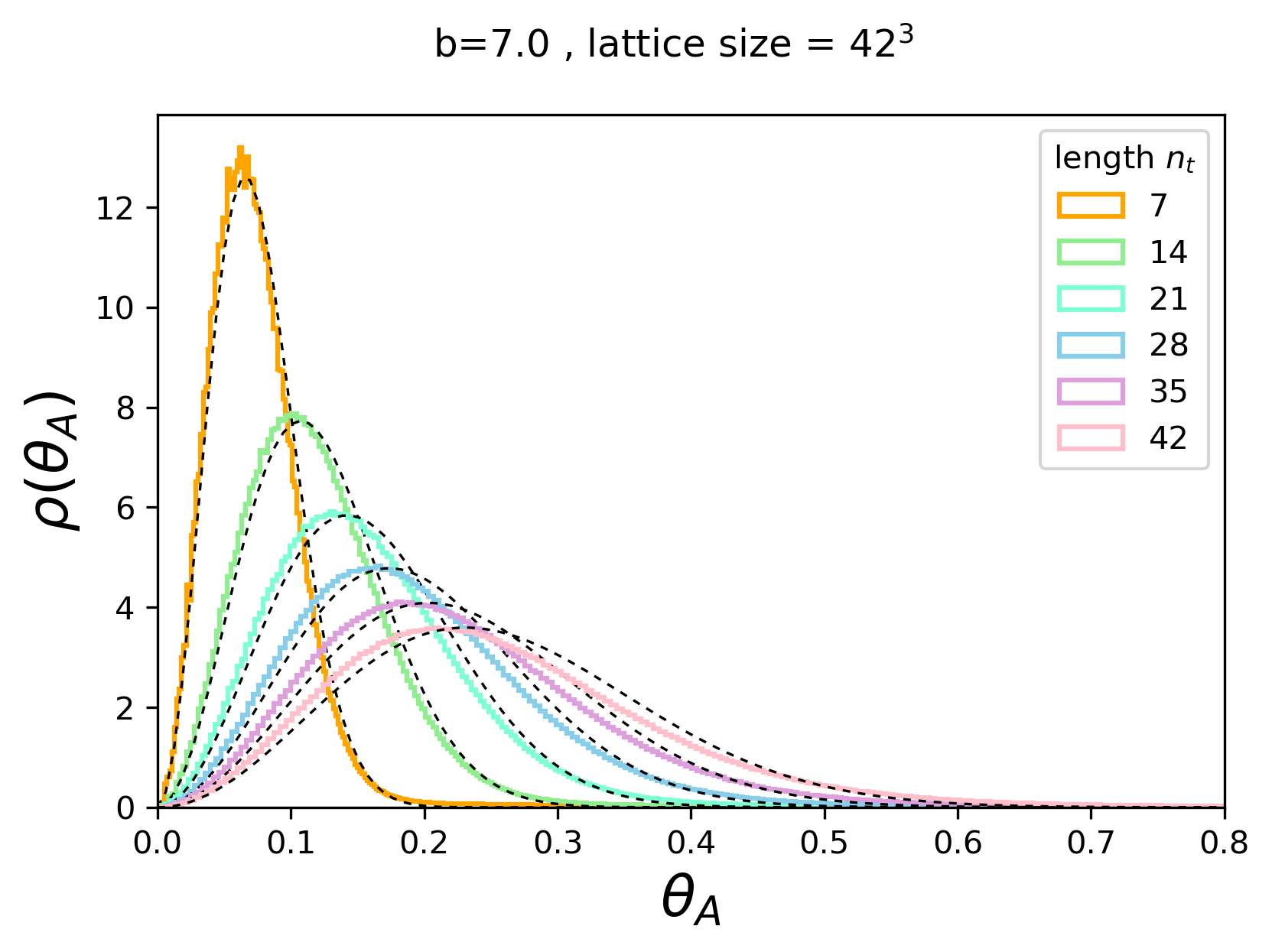}}
\scalebox{0.45}{
\includegraphics{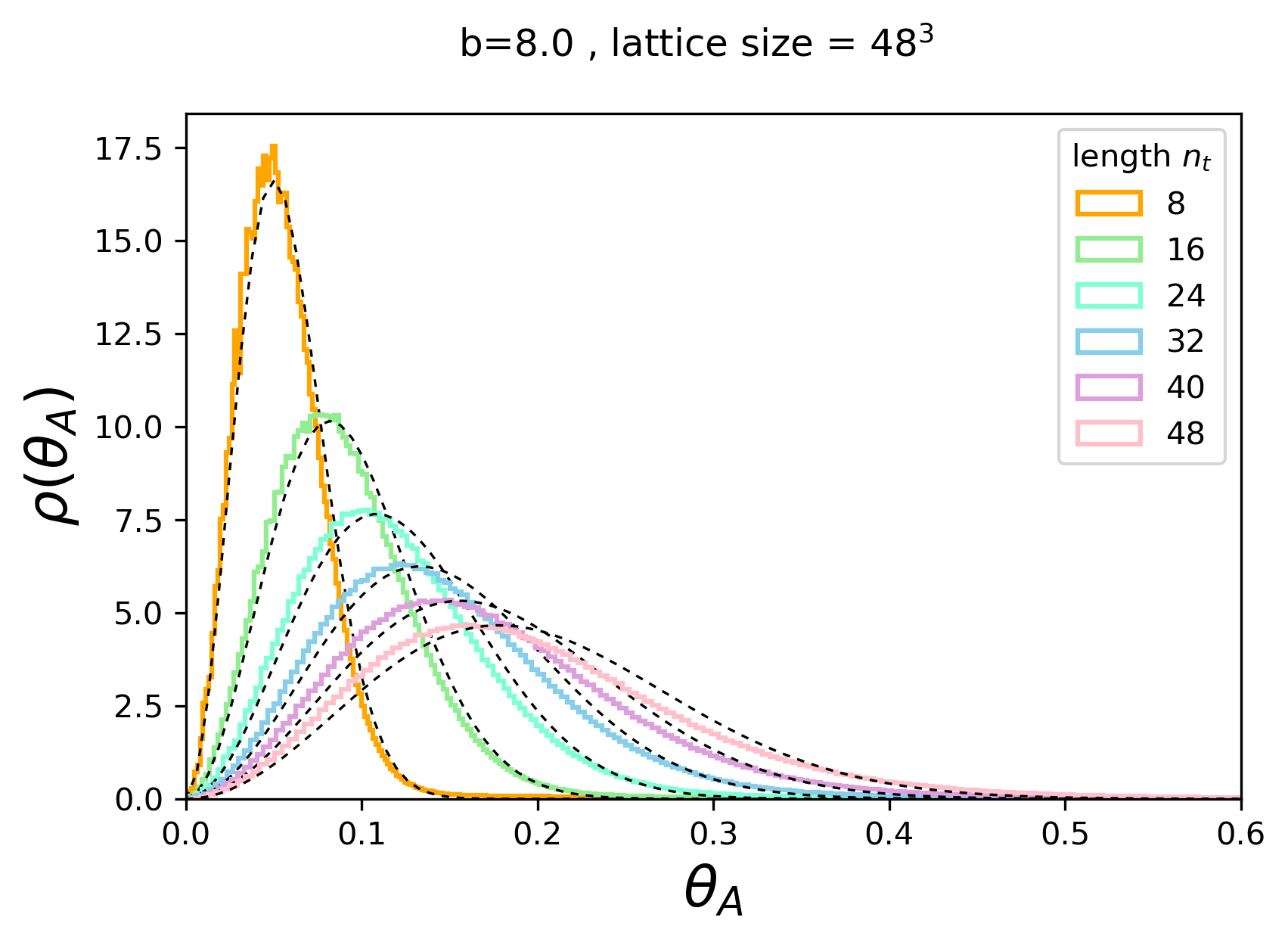}}
\end{center}
\caption{Comparison of GUE Wigner surmise rescaled by a factor of $\langle \theta_A \rangle$ (see eq.~\ref{GUEWig}) and distribution of $\theta_W$ from $W(L_t\times a)$  for inverse temperature $\beta=24$. There is slight deviation of $A(L_t, a)$ from the GUE Wigner distribution (shown above as dashed lines) which is prominent at smaller lattice spacing $a = \frac{4}{b}$ and larger $L_t$.  
}\label{fig:A-Wigner}
\end{figure}

Similarly, we also expect randomness of $A(L_t,a)$ and the corresponding phase $\theta_A$. In Fig.~\ref{fig:W-1*W-Wigner} and Fig.~\ref{fig:A-Wigner} we compare the distribution of $\theta_A$ to the GUE Wigner surmise. There is approximate agreement, although $\theta_A$ shows small but considerable deviations from the Wigner surmise at small $a$ (or equivalently, large $b$). $A$ is hence not precisely Gaussian random. However,  our claims regarding $A$ are not related to the precise form of the distribution, but only on the fact that it is randomly distributed.

In order to investigate the scaling of $\langle \theta_A \rangle$, we have taken $A$ loops at fixed physical length\footnote{
Because we define the physical length using the improved lattice spacing as $L_t=a_{\rm I}n_t$, and because $a_{\rm I}$ depends nontrivially on the coupling constant, we need to fit $\langle\theta_A\rangle$ as a function of $L_t$ so that we can see the $a_{\rm I}$ dependence at fixed $L_t$. See the bottom panel of Fig.~\ref{fig:width}. 
} and plotted them against the squared improved lattice spacing, $a_I^2$, as shown in the right panel of Fig.~\ref{fig:width}.
The ansatz $k_1 a_I^2 + k_2 a_I^3$ fits the data well, and near the origin we observe a scaling like $\langle \theta_A \rangle \propto a_I^2$. This scaling is consistent with the random walk picture: in the continuum limit we expect $\log A \sim a^2 \sqrt{L_t} \partial X$, and since $X$ is determined from a renormalized field, $\partial X$ should be of order $O(a_I^0)$. Therefore, we expect the scaling $\langle \theta_A \rangle \propto a_I^2$ at fixed physical length. %(Note that $\theta_A$ is the eigenvalues of $-i\log A$.)

\begin{figure}[htbp]
\begin{center}
\scalebox{0.6}{
\includegraphics{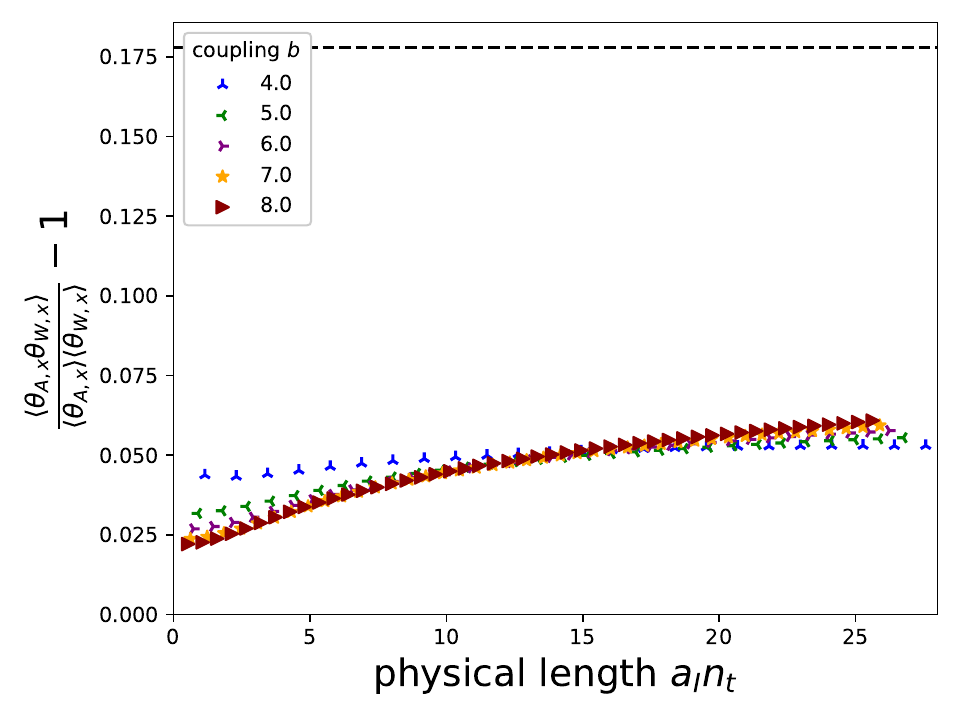}}
\end{center}
\caption{
$\frac{\langle\theta_{W,\vec{x}}\cdot\theta_{A,\vec{x}}\rangle}{\langle\theta_{W,\vec{x}}\rangle\cdot\langle\theta_{A,\vec{x}}\rangle}-1$ for $\beta=24$, plotted against the physical temporal length of the loops. As we can see, the correlation increases with $L_t$ but is always smaller than 0.178, shown by the dashed black line, which is the value obtained assuming maximal correlation. Numerical uncertainties are smaller than the size of the data points.
}\label{fig:AvsW Correlation}
\end{figure}

\begin{figure}[htbp]
\begin{center}
\scalebox{0.45}{
\includegraphics{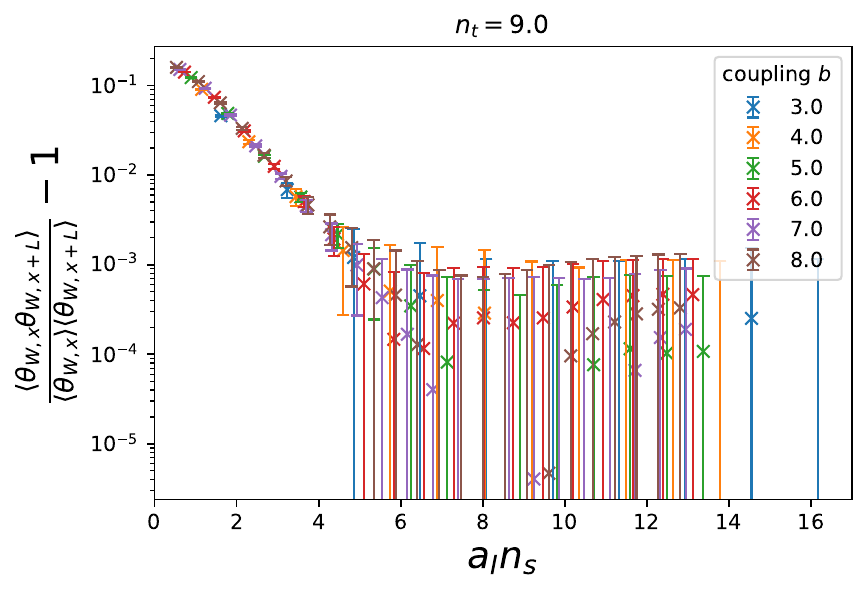}}
\scalebox{0.45}{
\includegraphics{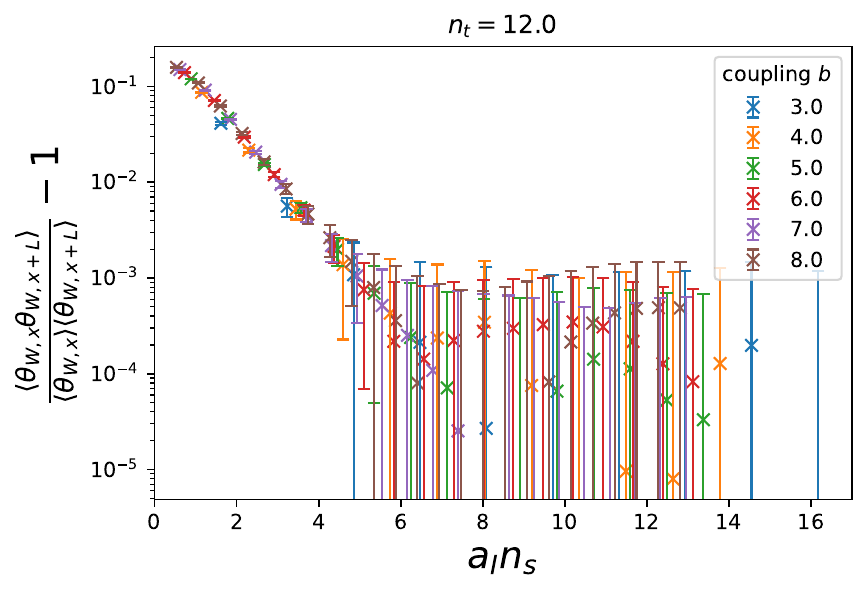}}
\scalebox{0.45}{
\includegraphics{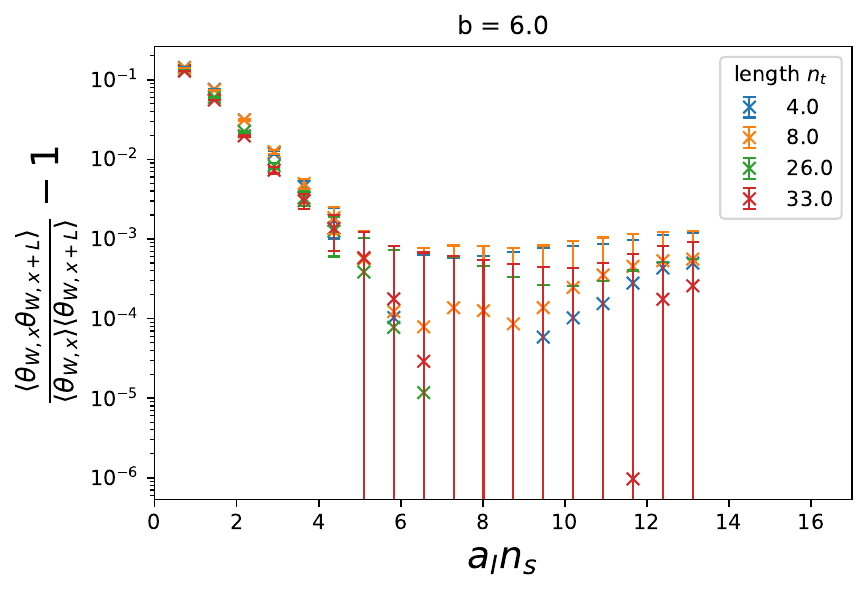}}
\scalebox{0.45}{
\includegraphics{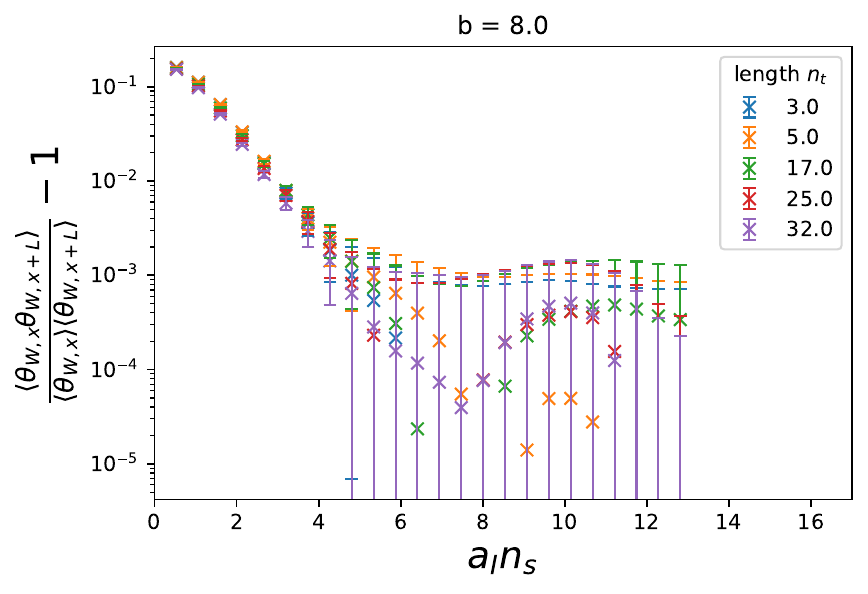}}
\end{center}
\caption{
$\frac{\langle\theta_{W,\vec{x}}\cdot\theta_{W,\vec{x}'}\rangle}{\langle\theta_{W,\vec{x}}\rangle\cdot\langle\theta_{W,\vec{x}'}\rangle}-1$ as a function of $|\vec{x}-\vec{x}'|=a_In_s$. Note that vertical axis is plotted in logarithmic scale.
}\label{fig:W-correlation}
\end{figure}

For the random-walk picture to be valid, $W$ and $A$, and hence $\theta_W$ and $\theta_A$, should have only a small correlation.
Therefore we also investigate this correlation, which is determined by
\begin{align}
\frac{\langle\theta_{W,\vec{x}}\cdot\theta_{A,\vec{x}}\rangle}{\langle\theta_{W,\vec{x}}\rangle\cdot\langle\theta_{A,\vec{x}}\rangle}-1\, . 
\end{align}
For completely uncorrelated data, this quantity must be zero. If the correlation is maximal and $\frac{\theta_{W,\vec{x}}}{\langle\theta_{W,\vec{x}}\rangle}=\frac{\theta_{A,\vec{x}}}{\langle\theta_{A,\vec{x}}\rangle}$ for any $\vec{x}$ and lattice configurations, then this quantity should be 
\begin{align}
\left(\frac{\langle\theta_{W,\vec{x}}\cdot\theta_{A,\vec{x}}\rangle}{\langle\theta_{W,\vec{x}}\rangle\cdot\langle\theta_{A,\vec{x}}\rangle}-1\right)_\text{max}=\frac{\langle\theta^2_{W,\vec{x}}\rangle }{\langle\theta_{W,\vec{x}}\rangle^2}-1=\int_0^\infty s^2\rho_\text{GUE}(s)\mathrm{d}s-1\simeq 0.178\, . 
\end{align}
 As we can see from Fig.~\ref{fig:AvsW Correlation}, the actual values are significantly smaller than 0.178, and hence $W$ and $A$ are not strongly correlated. Therefore, $W_{L_t}(\vec{x},\vec{x}+a\hat{\mu})$ and $W_{L_t}(\vec{x}',\vec{x}'+a\hat{\mu})$ should lose correlation if $\vec{x}$ and $\vec{x}'$ are sufficiently separated. Indeed, as shown in Fig.~\ref{fig:W-correlation}, 
 \begin{align}
\frac{\langle\theta_{W,\vec{x}}\cdot\theta_{W,\vec{x}'}\rangle}{\langle\theta_{W,\vec{x}}\rangle\cdot\langle\theta_{W,\vec{x}'}\rangle}-1
\to 0 
\end{align}
as $|\vec{x}-\vec{x}'|$ increases. Specifically, the decay is exponentially fast and the exponent is not sensitive to $L_t$. At $|\vec{x}-\vec{x}'|\gtrsim 4$, it is hard to distinguish it from zero numerically. 
 
In summary our observations are consistent with the random walk picture. The scaling of $\langle\theta_W\rangle\propto\sqrt{a^3n_t}=a\sqrt{L_t}$ and, at fixed physical length, $\langle\theta_A\rangle\propto a^2\sim a\cdot\langle\theta_W\rangle$ is also in accordance with these expectations. Eq.~\eqref{eq:def_ov_v} then implies that $v$ and its derivative is, like $X$, of order $a^0$. 
Moreover, as we could see in Fig.~\ref{fig:W-correlation}, $W_{L_t}(\vec{x},\vec{x}+a\hat{\mu})$ and $W_{L_t}(\vec{x}',\vec{x}'+a\hat{\mu})$ lose correlation if $\vec{x}$ and $\vec{x}'$ are sufficiently separated. Specifically, the correlation is negligible beyond the separation of order $\frac{1}{a}$ lattice units and order one physical distance.
The Casimir scaling follows from this random walk property. Although we refer to Ref.~\cite{Bergner:2023rpw} for detailed discussion, let us give a short, intuitive explanation here. Let us consider not-so-thin loops that are obtained as products of order $\frac{1}{a}$ number of $W$'s. These loops are almost uncorrelated, and hence, analogous to $W_j$'s in Sec.~\ref{sec:review}. Therefore, a random walk on the group manifold is essentially a random matrix product. 
When we take a product of order $\frac{1}{a}$ number of $W$'s, we can roughly estimate that the width of the Gaussian distribution $a\sqrt{L_t}$ adds up to $a\sqrt{L_t}\times\frac{1}{a}=\sqrt{L_t}$. To build an $L_t\times L_s$ loop, we need to take a product of $~L_s$ number of such loops, and hence, area law $\sim e^{-\sigma^2L_tL_s}$ follows. 
%%%%%%%%%%%%
%%%%%%%%%%%%
\section{Conclusion}\label{sec:conclusion}
\hspace{0.51cm}
%%%%%%%%%%%%
%%%%%%%%%%%%
In this paper, we studied three-dimensional SU(2) pure Yang-Mills to demonstrate the random walk property of the rectangular Wilson loop conjectured in Ref.~\cite{Bergner:2023rpw}, from which linear confinement with Casimir scaling follows. We confirmed the random-walk property that establishes the validity of the conjecture at least for this case of three-dimensional SU(2) Yang-Mills theory. We believe this is a generic feature of strongly coupled confining gauge theories. It is worth checking this for several other theories, most importantly for four-dimensional pure Yang-Mills and QCD. A small but nonzero deviation from Haar randomness should exist and it contains important information such as the spectrum of gauge-invariant excitations and string breaking~\cite{Hanada:2023krw,Hanada:2023rlk,Bergner:2023rpw}. It is important to understand the details of such a deviation in each theory.  

Random-matrix approaches of different kinds have been a powerful tool to study supersymmetric Wilson loops~\cite{Drukker:2000rr,Pestun:2007rz,Kapustin:2009kz,Drukker:2010nc}. It would be interesting if we could derive analogous statements from our findings. 

%%%%%%%%%%%%%%%%%%%%%%
%%%%%%%%%%%%%%%%%%%%%%
\begin{center}
\Large{\textbf{Acknowledgement}}
\end{center}
%%%%%%%%%%%%%%%%%%%%%%
%%%%%%%%%%%%%%%%%%%%%%
The authors would like to thank Nadav Drukker, Enrico Rinaldi, Hidehiko Shimada, and Hiromasa Watanabe for the discussions.
V.~G. thanks STFC for the Doctoral Training Programme funding (ST/W507854-2021 Maths DTP).
M.~H. thanks his STFC consolidated grant ST/X000656/1. 
G.~B.\ is funded by the Deutsche Forschungsgemeinschaft (DFG) under Grant No.~432299911 and 431842497.

\appendix
%%%%%%%%%%%%%%%%%%%%%
%%%%%%%%%%%%%%%%%%%%%
\section{Why Polyakov line random-walks}\label{sec:Hanada-Watanabe-review}
\hspace{0.51cm}
%%%%%%%%%%%%%%%%%%%%%
%%%%%%%%%%%%%%%%%%%%%
In this appendix, we explain the origin of the slowly varying Haar randomness of the Polyakov line in the confined phase~\cite{Hanada:2023rlk,Bergner:2023rpw}. Although the same logic applies to various other theories, we focus on  $(1+3)$-dimensional SU($N$) pure Yang-Mills theory.

We work on the Hamiltonian formulation and use the coordinate basis. As a concrete setup, we use the Kogut-Susskind formulation. There are operators $\hat{U}_{j,\vec{n}}$ ($j=1,2,3=x,y,z$) describing the unitary link variable $U_{j,\vec{n}}$ on a link connecting the site $\vec{n}$ and $\vec{n}+\hat{j}$. It acts on the coordinate eigenstate $\ket{g}$ ($g\in\mathrm{SU}(N)$) as $\hat{U}_{j,\vec{n}}\ket{g}_{j,\vec{n}}=g\ket{g}_{j,\vec{n}}$. The Hilbert space is $\mathcal{H}_{\rm ext}=\otimes_{j,\vec{n}}\mathcal{H}_{j,\vec{n}}$, where $\mathcal{H}_{j,\vec{n}}=\mathrm{Span}\{\ket{g}_{j,\vec{n}}|g\in\mathrm{SU}(N)\}$. The subscript `ext' stands for the \textit{extended} Hilbert space, meaning that $\mathcal{H}_{\rm ext}$ contains SU($N$) non-singlets. We can also obtain the gauge-invariant Hilbert space $\mathcal{H}_{\rm inv}$ by collecting SU($N$)-singlet states. 
Let us use $\hat{\Omega}$ to denote the local SU($N$) transformation $\ket{g}_{j,\vec{n}}\to\ket{\Omega_{\vec{n}}g\Omega^{-1}_{\vec{n}+\hat{j}}}_{j,\vec{n}}$. We use $\mathcal{G}$ to denote the set of all local SU($N$) transformations. Then, the canonical partition function at temperature $T=\beta^{-1}$ is~\cite{Hanada:2020uvt}
\begin{align}
    Z(\beta) = \mathrm{Tr}_{\mathcal{H}_{\text{inv}}} e^{-\beta \hat{H}} = \frac{1}{\text{vol}\mathcal{G}}\int_{\mathcal{G}} d\Omega \mathrm{Tr}_{\mathcal{H}_{\text{ext}}}\left(\hat{\Omega} e^{-\beta \hat{H}} \right)\; .
\end{align}
By inserting $\Omega$ and integrating it over $\mathcal{G}$, a projection to the singlet state is realized. In terms of the path integral, $\hat{\Omega}$ is the Polyakov loop~\cite{Hanada:2020uvt}. Therefore, we use $P$ instead of $\Omega$ in the following. 

An energy eigenstate $\ket{\Phi}\in\mathcal{H}_{\text{ext}}$ with energy $E_\Phi$ contributes to the partition function as
\begin{align}
    \frac{e^{-\beta E_\Phi}}{\text{vol}\mathcal{G}}
    \int_{\mathcal{G}} dP
    \bra{\Phi}\hat{P}\ket{\Phi}\, .
    \label{eq:enhancement_factor}
\end{align}
We can see that such a Polyakov loop $P$ that does not change the state $\hat{\Phi}$ too much, and hence $\ket{\Phi}$ and $\hat{P}\ket{\Phi}$ have a large overlap, contributing significantly to the partition function. 

The confined ground state in or sufficiently close to the continuum limit is a wave packet localized around $U_{j,\vec{n}} = \mathbf{1}$ up to gauge transformation. Any wave packets connected to this wave packet by a gauge transformation are equivalent. Therefore, we need to consider a wave packet around 
\begin{align}
U_{j,\vec{n}}=g_{\vec{n}}^{-1}g_{\vec{n}+\hat{j}}\, , 
\qquad g\in\mathcal{G}\, . 
\end{align}
Let us consider the action of the `constant' Polyakov loop $P_{\vec{n}} \equiv g^{-1}_{\vec{n}} V g_{\vec{n}}$  ($V\in\mathrm{SU}(N)$) on such a vacuum configuration. Here, we assume $V$ is independent of the lattice site $\vec{n}$, and hence, $P_{\vec{n}}$ is constant up to the local SU($N$) transformation. Regardless of the choice of $V$, the wave packet under consideration does not move because 
\begin{align}
    P_{\vec{n}}^{-1} 
    (g_{\vec{n}}^{-1}g_{\vec{n}+\hat{j}}) 
    P_{\vec{n}+\hat{j}} 
    = 
    (g_{\vec{n}}^{-1} V^{-1} g_{\vec{n}}) 
    (g_{\vec{n}}^{-1} g_{\vec{n}+\hat{j}}) 
    (g_{\vec{n}+ \hat{j}}^{-1} V g_{\vec{n}+\hat{j}}) 
    = 
    g_{\vec{n}}^{-1} g_{\vec{n}+\hat{j}}\, . 
\end{align}
Therefore, any $V\in\mathrm{SU}(N)$ contributes equally to the partition function. 
We can also consider a `local' $SU(N)$ transformation with $V$ now being dependent on the site $\vec{n}$ with the Polyakov line taking the form  $P_{\vec{n}} \equiv \Omega^{-1}_{\vec{n}} V_{\vec{n}} \Omega_{\vec{n}}$. 
In this case, we have 
\begin{align}
    P_{\vec{n}}^{-1} 
    (g_{\vec{n}}^{-1}g_{\vec{n}+\hat{j}}) 
    P_{\vec{n}+\hat{j}}
    = 
    (\Omega_{\vec{n}}^{-1} V_{\vec{n}}^{-1} \Omega_{\vec{n}})
    (\Omega_{\vec{n}}^{-1} \Omega_{\vec{n}+\hat{j}}) 
    (\Omega_{\vec{n}+ \hat{j}}^{-1} V_{\vec{n}+\hat{j}} \Omega_{\vec{n}+\hat{j}})
    =
    \Omega_{\vec{n}}^{-1} V_{\vec{n}}^{-1} V_{\vec{n}+\hat{j}} \Omega_{\vec{n}+\hat{j}}\, .. 
\end{align}
If $V_{\vec{n}}$ is slowly varying, then $V_{\vec{n}}^{-1} V_{\vec{n}+\hat{j}}$ is close to $\mathbf{1}$, and hence the contribution to the partition function is large. Because there is no constraint other than this slowly varying nature, slowly varying Haar randomness follows. 
%%%%%%%%%%%%%%%%%%%%%%%%%
%\bibliographystyle{unsrt}
\bibliographystyle{utphys}
\bibliography{RMP-Confinement}

\end{document}